\documentclass[pre,twocolumn,amsmath,amssymb,showpacs]{revtex4}
\usepackage{graphicx}
\usepackage{dcolumn}
\usepackage{bm}
\begin{document}
%
%
\newcommand{\bfGamma}{\mbox{\boldmath $\it\Gamma$}}
\newcommand{\bfsigma}{\mbox{\boldmath $\it\sigma$}}
\newcommand{\bfp}{\mbox{\boldmath $\it p$}}
\newcommand{\bfq}{\mbox{\boldmath $\it q$}}
\newcommand{\bfe}{\mbox{\boldmath $\it e$}}
\newcommand{\calD}{\mathcal{D}}
\newcommand{\vspfigA}{\vspace{0.45cm}
   \rule{85mm}{0.3mm}
   \vspace{-0.15cm}}  
\newcommand{\vspfigB}{\vspace{-0.05cm}
   \rule{85mm}{0.3mm}
   \vspace{0.35cm}} 
\newcommand{\widthfigA}{0.41\textwidth}
\newcommand{\widthfigB}{0.49\textwidth} 
\newcommand{\ls}{\hspace{0.5cm}}
\newcommand{\sumindex}[2]{
   \substack{#1 \\ (#2)} }
%
%
%
%
\title{
   Hopping dynamics for localized Lyapunov vectors 
   in many-hard-disk systems }
\author{Tooru Taniguchi and Gary P. Morriss}
\affiliation{School of Physics, University of New South Wales, 
Sydney, New South Wales 2052, Australia}
\date{\today}
\begin{abstract}
   The dynamics of the localized region of the Lyapunov vector 
for the largest Lyapunov exponent is discussed 
in quasi-one-dimensional hard-disk systems at low density. 
   We introduce a hopping rate to quantitatively describe the 
movement of the localized region of this 
Lyapunov vector, and show that it is a decreasing function of 
hopping distance, 
implying spatial correlation of the localized regions.  
   This behavior is explained quantitatively 
by a brick accumulation model derived from 
hard-disk dynamics in the low density limit, 
in which hopping of the localized Lyapunov vector is 
represented as the movement of the highest brick position. 
   We also give an analytical expression for the 
hopping rate, which is obtained us a sum 
of probability distributions for brick height configurations 
between two separated highest brick sites. 
   The results of these simple models are in good agreement with 
the simulation results for hard-disk systems.
\end{abstract}
%
\vspace{1cm}
\maketitle

\section{Introduction}

   
   The dynamical instability and many-body nature 
play essential roles in the justification 
of a statistical treatment for deterministic dynamical systems. 
   The dynamical instability is described as a rapid expansion  
of the difference between two nearby trajectories, 
namely the Lyapunov vector, and the system is called chaotic 
if at least one exponential rate (Lyapunov exponent) 
of divergence or contraction of the amplitude 
of Lyapunov vector is positive. 
   The Lyapunov exponent $\lambda$ is defined for each 
independent direction of the phase space, 
so the chaotic properties of many-body systems can be  
characterized by an ordered set of Lyapunov exponents, 
the so-called Lyapunov spectrum 
$\{\lambda^{(1)},\lambda^{(2)},\cdots\}$ 
where  
$\lambda^{(1)}\geq \lambda^{(2)}\geq\cdots$. 
   The Lyapunov spectrum  
is connected to the contraction rate of the phase 
space volume (roughly speaking, the dissipation rate)  
through the sum of all the Lyapunov exponents, 
allowing the   
calculation of transport coefficients 
from the Lyapunov spectra \cite{Eva90a,Gas98,Dor99}. 
   The conjugate pairing rule for Lyapunov spectra  
reduces the calculation of the 
sum of all the Lyapunov exponents to 
the sum of just one pair of Lyapunov exponents 
\cite{Dre88,Eva90b,Det96a,Tan02a}. 
   The set of all positive Lyapunov exponents specifies  
the natural invariant measure \cite{Eck85,Eck86}, 
which is used to calculate various quantities 
using periodic orbit theory \cite{Cvi05} and led to 
the first form of the fluctuation theorem 
\cite{Eva93,Gal95,Eva02}. 
   On the other hand, recently, much attention has 
been paid to individual Lyapunov exponents 
and their Lyapunov vectors 
for many-body systems.
   As each Lyapunov exponent has the dimensions of
inverse time, the Lyapunov 
spectrum can be regarded as a time-scale spectrum.  
   From this point of view, Lyapunov exponents with small 
absolute values are connected to large and macroscopic 
time-scale behavior of many-body systems, and 
in this region the wave-like structure of Lyapunov vectors,  
known as the Lyapunov modes, is observed 
\cite{Pos00,Tan02c,Wij04,Yan04,Tan03a}. 
   The Lyapunov mode is a reflection of a collective movement 
(phonon mode) of many-body systems, and comes from dynamical 
conservation laws and translational invariance 
\cite{Tan03a,Tan04b,Eck05,Tan05a}. 
   On the other hand, large Lyapunov exponents are 
dominated by small and microscopic time-scale movement, 
and in this region the spatially localized behavior of 
Lyapunov vector, the so-called Lyapunov localization, appears 
\cite{Man85,Kan86,Liv89,Fal91,Pik01,Mil02,Tan03b}. 
   For the largest Lyapunov exponent 
of many-body chaotic systems, analytical calculations  
have been attempted 
\cite{Cas96,Cas00,Zon98,Bei00,Zon02}. 
   The time-scale separation in many-body systems 
is crucial to extract a macroscopic 
dynamics from microscopic many-body dynamics, 
and the Lyapunov spectrum allows us to discuss it 
dynamically. 

   As one of the features of Lyapunov vectors 
for many-body systems, the Lyapunov localization 
appears as a behavior in which the Lyapunov vector components 
for a few particles are significantly larger than the other 
components. 
   Moreover, the localized region of a Lyapunov vector moves  
as a function of time. 
   The magnitude of the localization of each Lyapunov vector 
can be measured quantitatively by 
the Lyapunov localization spectrum, which is defined 
as a set of exponential functions of entropy-like 
quantities for the normalized amplitudes of the Lyapunov vector 
components \cite{Tan03b,Tan05b,For05}. 
   We have previously reported that the Lyapunov 
localization spectra show a bending behavior 
at low density, and its connection with  
kinetic theory properties (e.g. the Krylov relation 
for the largest Lyapunov exponent, and  
the mean free time being inversely-proportional to density, etc.) 
is discussed for many-hard-disk systems  \cite{Tan03b,Tan05b}. 
   However, the Lyapunov localization spectrum 
requires taking a time-average and characterizes only the  
static localization of the Lyapunov vectors. 
   The physical meaning of the movement of the 
localized regions of Lyapunov vectors is not clearly understood.  

   The principal aim of this paper is to discuss 
dynamically the movement of the localized region 
of the Lyapunov vectors in many-hard-disk systems 
at low density. 
   To discuss this problem we use the fact 
shown in Ref. \cite{Tan03b} that at low density 
only two particle components of the normalized Lyapunov vector 
for large Lyapunov exponents have a non-zero value.  
   This is due to the short-range of 
particle interactions in a many-hard-disk system. 
   The movement of the localized region 
of the Lyapunov vector appears 
to be a series of jumps or hops, so we can introduce 
a hopping rate to describe the dynamics. 
   To simplify the problem, in this paper 
we consider quasi-one-dimensional systems, 
in which the system width is so narrow that disks 
always remain in the same order \cite{Tan03a,Tan05a,Tan03b,Tan05b}. 
   We show that this hopping rate depends 
on the hopping distance, 
and is a decreasing function of 
the hopping distance. 
   This implies that there is a spatial correlation 
among localized regions of Lyapunov vectors. 
   
   We explain the hopping-distance dependence 
of the hopping rate in many-hard-disk systems 
in two ways. 
   In the first approach we use a simple model expressed 
as an accumulation of bricks. 
   Here, the hopping of the localized region of the 
Lyapunov vectors is expressed as a change in the 
position of the highest brick site. 
   This model is a one-dimensional version of 
the so-called clock model, which has been used to calculate 
Lyapunov exponents for many-hard-disk systems 
\cite{Zon98,Bei00,Zon02}. 
    In this paper we demonstrate that 
this model can reproduce the largest 
Lyapunov exponent for quasi-one-dimensional 
hard-disk systems. 
   As the second approach to the hopping behavior 
of the localized Lyapunov vectors, we propose 
an analytical method to calculate the hopping rate 
from the sum of probability 
distributions for the brick height configurations 
between two separated highest brick sites. 
   Using this analytical approach  
we can also discuss the relation between the hopping rate 
and the Lyapunov exponent. 
   Hopping rates calculated by these two approaches 
are in good agreement with the ones
for quasi-one-dimensional hard-disk systems.

   The outline of this paper is as follows. 
   In Sec. \ref{QuasiOneDimensionalSystem} we introduce 
the quasi-one-dimensional hard-disk system, and 
show the localized behavior of the Lyapunov vector 
corresponding to the largest Lyapunov exponent at low density.  
   In Sec. \ref{HoppingRate} we introduce the hopping rate 
of the localized region of the Lyapunov vectors, and show 
the hopping-distance dependence for many-hard-disk systems. 
   In Sec. \ref{brickmodel} we discuss the hopping rate using 
a brick accumulation model. 
   In Sec. \ref{AnalyticalExpression} we propose an 
analytical expression for the hopping rate. 
   Section \ref{ConclusionRemarks} is our conclusion and 
some remarks. 
   In Appendix \ref{HoppingRateLocalizedLyapunovVector} 
we discuss the technical details of the calculation 
of the hopping rate. 
   In Appendix \ref{ClockModelManyHardParticleSystems} 
we give a microscopic derivation of the brick accumulation model.

\section{Quasi-One-Dimensional System 
and Localization of Lyapunov Vectors}
\label{QuasiOneDimensionalSystem}

   The system which we consider in this paper 
is a quasi-one-dimensional system 
consisting of $N$ hard-disks  
in periodic boundary conditions. 
   All of the particles are identical with radius $R$ 
and mass $M$, and the shape of the system is rectangular 
with the length $L_{x}$ and the width $L_{y}$ 
satisfying the inequality $2R < L_{y} < 4R$. 
   The schematic illustration of the system is given in Fig. 
\ref{figA1qua1dim}, in which we number particles 
$1,2,\cdots,N$ from the left to right in this system. 
   For the actual numerical results shown in this paper, 
we used:  
the radius of a particle $R=1$, 
the mass of a particle $M=1$, 
the total energy of the system $E=N$, 
the system width $L_{y}=2R(1+10^{-6})$, and 
the system length $L_{x}=NL_{y}(1+d)$ 
with the constant $d$ controlling the density $\rho \equiv 
N\pi R^{2}/(L_{x}L_{y})$.  
   In the quasi-one-dimensional system, the particle interactions 
are restricted to nearest-neighbor particles only, so  
particles remain in the same order. 
   These features require less calculation effort and 
a simpler representation of results for  
quasi-one-dimensional system compared with 
fully two-dimensional systems. 
   The quasi-one-dimensional system has already been used to 
investigate the localized behavior of Lyapunov vectors 
\cite{Tan03b,Tan05b},  
the wave-like structure of Lyapunov vectors 
\cite{Tan03a,Tan05a,Tan04b} 
and the transition between quasi-one-dimensional  
and fully two-dimensional systems \cite{For04}. 
%
\begin{figure}[t!] 
\vspfigA
\caption{
      A quasi-one-dimensional system consisting of 
   hard-disks with radius $R$. 
      The length $L_{x}$ and  
   the width $L_{y}$ of the system control the density. 
      The width $L_{y}$ satisfies the inequality 
   $2R < L_{y} < 4R$, so that the disks remain in the same order 
   and particles can be numbered 
   $1,2,\cdots,N$ from the left to right. 
   }
\label{figA1qua1dim}
\vspfigB\end{figure}  

   The dynamics of Lyapunov vectors in many-hard-disk systems 
is well established, and readers should refer to the references, 
for example, Ref. \cite{Del96},   
for more detailed discussions. 
   In many-hard-disk systems 
the dynamics is separated into a free-flight part and 
a collision part, and the free-flight part of the dynamics 
is integrable. 
   This property allow us to express the dynamical evolution  
as a simple multiplication of time-evolutional matrices 
for the free-flight dynamics and the collision dynamics, 
leading to a fast and more accurate numerical simulation 
than for soft-core interaction models. 
   For numerical calculations of the Lyapunov vectors 
shown in this paper, we use the algorithm developed by 
G. Benettin, \textit{et al} \cite{Ben76} and 
I. Shimada and T. Nagashima \cite{Shi79} 
(also see Refs. \cite{Ben80a,Ben80b}). 
   This algorithm is characterized by intermittent 
(e.g. after every collision) re-orthogonalization 
and renormalization of Lyapunov vectors, 
preventing a divergence of the amplitude of the  
Lyapunov vectors.

   In this paper we use the notation 
$\delta\bfGamma^{(n)}(t) \equiv (\delta\bfGamma_{1}^{(n)}(t), 
\delta\bfGamma_{2}^{(n)}(t),\cdots,\delta\bfGamma_{N}^{(n)}(t))$ 
for the Lyapunov vector 
corresponding to the $n$-th Lyapunov exponent $\lambda^{(n)}$
at time $t$. 
   Here, $\delta\bfGamma_{j}^{(n)}(t)$ is the Lyapunov vector 
component contributed by the $j$-th particle 
in the $n$-th Lyapunov exponent at time $t$.
   To express the localized behavior of the Lyapunov vectors 
we introduce the quantity $\gamma_{j}^{(n)}(t)$ as 
\begin{eqnarray}
   \gamma_{j}^{(n)}(t) \equiv 
   \frac{\left|\delta\bfGamma_{j}^{(n)}(t)\right|^{2}}
   {\sum\limits_{k=1}^{N}
   \left|\delta\bfGamma_{k}^{(n)}(t)\right|^{2}},  
\label{gamma}\end{eqnarray}
which is the normalized amplitude of the 
Lyapunov vector component for the $j$-th particle 
for the $n$-th Lyapunov exponent $\lambda^{(n)}$ 
at time $t$. 
   The localized behavior of Lyapunov vector, 
namely the Lyapunov localization, 
is the phenomenon where  
only a few of the $\gamma_{j}^{(n)}(t)$, $j=1,2,\cdots,N$ 
have a non-zero value at any time $t$.

\begin{figure}[t!]
\vspfigA
\caption{
      The normalized amplitude $\gamma_{j}^{(1)}$ of the 
   Lyapunov vector particle component  
   corresponding to the largest Lyapunov exponent 
   as a function of the collision number $n_{t}$ 
   and the particle index $j$ in a quasi-one-dimensional 
   hard-disk system with $d=10^{5}$ 
   and $N=50$. 
      On the base of this graph is a contour plot of  
   $\gamma_{j}^{(1)}$ at the level $0.2$. 
   }
\label{figA2loczAmpDyn}
\vspfigB\end{figure}  
%
   Fig. \ref{figA2loczAmpDyn} is an example of 
the Lyapunov localization. 
   It is a graph of the normalized amplitude 
$\gamma_{j}^{(1)}$ of the Lyapunov vector components 
for the $j$-th particle corresponding to 
the largest Lyapunov exponent $\lambda^{(1)}$
as a function of the particle index $j$ 
and the collision number $n_{t}$ 
($\approx t/\tau$ with time $t$ and the mean free time $\tau$). 
   The system is a quasi-one-dimensional system 
consisting of $50$ hard-disks and with $d=10^{5}$ (the density 
$\rho\approx 7.85\times 10^{-6}$). 
   From this figure we recognize that the non-zero components 
of the Lyapunov vector are concentrated at two 
nearest-neighbor particles. 
   This characteristic is shown for hard-disk systems 
at low density and for large Lyapunov exponents 
in general, using the Lyapunov localization 
spectrum \cite{Tan03b,Tan05b}. 
   Moreover, we observe that such localized regions of 
Lyapunov vector components 
move with time in discrete jumps or hops, 
in Fig. \ref{figA2loczAmpDyn}.  
   This characteristic hopping has already been 
shown in Ref. \cite{Tan03b} and 
that such spatial hopping of the non-zero components 
of $\gamma_{j}^{(n)}(t)$ 
are caused by some particle-particle collisions. 
   However, not every collision causes  
a hopping of the localized Lyapunov vector.  
   In this sense, particle collisions themselves 
are not sufficient 
to explain the hopping movement of the 
Lyapunov localization.  

   The Lyapunov localization appears in the Lyapunov vectors 
corresponding to large Lyapunov exponents in general, but 
in this paper for simplicity 
we consider only the Lyapunov vector 
$\delta\bfGamma^{(1)}$ corresponding to 
the largest Lyapunov exponent $\lambda^{(1)}$.

\section{Hopping Rate of Localized Lyapunov Vectors}
\label{HoppingRate}

   An advantage of the quasi-one-dimensional system 
in investigations of Lyapunov localization is that the 
movement of particles in the transverse direction
are suppressed, and 
roughly speaking, the particle sequence corresponds  
to the particle's position.  
   Noting this feature, in this section 
we describe the hopping behavior 
of spatially localized Lyapunov vectors as the 
hopping of particle indices whose $\gamma_{j}^{(n)}$ 
defined by Eq. (\ref{gamma}) have non-zero values. 

   As shown in Refs. \cite{Tan03b,Tan05b} the 
particle indices with non-zero $\gamma_{j}^{(1)}$ 
are a pair of nearest-neighbor particles in the low density limit, 
and we can introduce the hopping distance $h$ at each collision 
as the change of particle indices. 
   In this definition the hopping distance 
$h$ is an integer number satisfying  
the inequality $-[N/2] \leq h \leq [N/2]$,  
where $[x]$ is the integer part of the real number $x$.  
   It should also be noted that we take 
the particle index $j$ as the one equivalent to the index $j\pm N$, 
because of periodic boundary conditions, so the hopping distance 
$h\pm N$ is equivalent to $h$. 
   More technical details of the calculation of the 
hopping distance $h$ are given in Appendix 
\ref{HoppingRateLocalizedLyapunovVector}. 
   Using this hopping distance $h$, in numerical simulations,   
we count the number $N_{T}(h)$ of hops with hopping distance 
$h$ in a time-interval $T$, and we define 
the hopping rate $P_{N} (h)$ as a function 
proportional to $N_{T}(h)$ as $T\rightarrow\infty$. 
   In this paper we use the hopping rate  
$P_{N} (h)/P_{N} (1) = \lim_{T\rightarrow\infty}N_{T}(h)/N_{T}(1)$ 
normalized by $P_{N} (1)$. 
   From the reflection symmetry of the quasi-one-dimensional system 
in the longitudinal direction, the 
hopping rate $P_{N} (h)$ must be symmetric, 
namely $P_{N} (h) = P_{N} (-h)$.  
   We use this hopping rate to quantitatively
discuss the hopping dynamics for localized Lyapunov vectors.

\begin{figure}[t!]
\vspfigA
\caption{
      The normalized hopping rates $P_{N} (h)/P_{N} (1)$ 
   for quasi-one-dimensional systems of different sizes
   at the same density  ($d=10^{3}$)  
   as a function of the hopping distance $|h|$. 
      The number of particles is   
   $N=25$ (circles), 
   $50$ (triangles), and $75$ (squares).  
      The graphs are log-log plots with error bars are given 
   by  $|P_{N} (h)-P_{N} (-h)|/P_{N} (1)$.  
      The lines are fits of the numerical data 
   to functions based on Eq. (\ref{normalizedP}) 
   with power functions 
   as the asymptotic form of the hopping rate. 
   }
\label{figB1hopRateN}
\vspfigB\end{figure}  
%
   Figure \ref{figB1hopRateN} shows log-log plots of 
the hopping rates 
$P_{N} (h)/P_{N} (1)$ normalized by $P_{N} (1)$ 
as a function of $|h|$ 
in a quasi-one-dimensional hard-disk 
system with $d=10^{3}$ at different numbers of particles; 
$N=25$ (circles), $50$ (triangles), and $75$ (squares).  
   Noting the symmetric property $P_{N} (-h)=P_{N} (h)$, 
we use $|P_{N} (h)-P_{N} (-h)|/P_{N} (1)$ as error bars 
in Fig. \ref{figB1hopRateN}. 
   It is clear from Fig. \ref{figB1hopRateN} that 
the hopping rate $P_{N} (|h|)$ decreases as $|h|$ increases. 
   This implies that there is 
a spatial correlation among the localized 
regions of a Lyapunov vector, rather than a random hopping. 

   The {\it turn-up} in the tail of the normalized hopping rate 
$P_{N} (h)/P_{N} (1)$ in Fig. \ref{figB1hopRateN} can be 
explained as an effect of periodic boundary conditions. 
   Under periodic boundary conditions the hopping distance 
$h + jN$ 
for any integer $j$ is observed as the hopping distance 
$h$ in the quasi-one-dimensional system consisting 
of $N$ hard disks. 
   Therefore, using the hopping rate $P_{\infty}(h)$ 
for the thermodynamics limit ($N\rightarrow\infty$ at  
a fixed density) for $h=0,\pm 1,\pm 2,\cdots \pm\infty$, 
the hopping rate $P_{N} (h)$ for a finite system 
should be represented as 
\begin{eqnarray}
   P_{N}(h) =  \sum_{j=-\infty}^{\infty}  P_{\infty}(h+jN)
\label{normalizedP}\end{eqnarray}
for $h=-[N/2], -[N/2]+1,\cdots, [N/2]$.
   The terms on the right-hand side of Eq. (\ref{normalizedP})  
with $j \neq 0$ cause the {\it turn-up} in the tail 
of the hopping rate $P_{N}(h)$ for a finite system. 
   In this sense we can explain the $N$-dependence of 
the hopping rate $P_{N} (h)$ using an $N$-independent 
asymptotic distribution $P_{\infty}(h)$. 
   To make this explanation convincing we fitted the numerical data 
for $P_{N} (h)/P_{N} (1)$ shown in Fig. \ref{figB1hopRateN} 
to the function 
$f(h)\equiv\alpha [h^{\beta} +\sum_{j=1}^{2} 
(|h+jN|^{\beta}+|h-jN|^{\beta})]$
assuming that the decay of $P_{\infty}(h) = \alpha |h|^{\beta}$  
with fitting parameters $\alpha$ and $\beta$, 
and neglecting the higher order small terms for $|j|\geq 3$. 
   Here, the fitting lines are 
dotted for $N=25$, the broken for $N=50$ and 
solid for $N=75$. 
\begin{table}[t!] 
\caption{
      Fitting parameters for the power function 
   $P_{\infty} (h) = \alpha |h|^{\beta}$ 
   for quasi-one dimensional systems of different numbers of 
   particles $N$ with $d=10^{3}$. 
      Notice that the coefficients $\alpha$ and
   $\beta$ are essentially independent of $N$.
    }
\label{Ndepenalphabeta}
\vspfigA
\begin{tabular}{c|cc}
\hline
\hline
   \makebox[5em]{$N$} & 
   \makebox[9em]{$\alpha $} & 
   \makebox[8em]{$\beta $}  \\
   \hline
   25 & 0.718 & 1.67 \\
   50 & 0.788 & 1.72 \\
   75 & 0.753 & 1.71 \\
   \hline \hline
\end{tabular}
\vspfigB
\end{table}%
   Note that these sets $(\alpha,\beta)$ 
of fitting parameter values, summarized in Table 
\ref{Ndepenalphabeta}, are almost independent 
of the number of particles $N$. 
   As shown in Fig. \ref{figB1hopRateN},  
these fitting lines nicely reproduce the values of the 
numerical data, especially the {\it turn-ups}.  
    
   We can also calculate the hopping rate $P_{N} (h)$ at 
different densities, as long as the density is low enough 
so that only a few of the normalized Lyapunov vector 
component amplitudes 
$\{\gamma_{j}^{(n)}\}_{j}$ have non-zero values. 
   In Appendix \ref{HoppingRateLocalizedLyapunovVector} 
we show that the normalized hopping rate $P_{N} (h)/P_{N} (1)$ is 
almost density-independent at least for $10^{3}<d<10^{5}$, 
(namely in the density region 
$7.85\times 10^{-6}<\rho< 7.85\times 10^{-4}$).
   However, a subtle increase of the normalized hopping rate 
as the density decreases is recognizable in this low density region.

\section{Brick Accumulation Model}
\label{brickmodel}

   As shown in Sec. \ref{HoppingRate} the hopping rate of the 
localized region of the Lyapunov vector is correlated spatially. 
   In this section we explain this characteristic using 
a simple one-dimensional model, 
which we call the \emph{brick accumulation model}.

\begin{figure}[t!]
\vspfigA
\caption{
      The brick accumulation model expressing the dynamics of 
   the amplitude 
   of the Lyapunov vector component for each particle 
   in a quasi-one-dimensional system. 
   }
\label{figC1brickModel}
\vspfigB\end{figure}  
%
   A schematic illustration of the brick accumulation model, 
or simply the brick model, 
is given in Fig. \ref{figC1brickModel}. 
   It is a one-dimensional model with $N$ sites in 
the horizontal direction. 
   The bricks are dropped at random and occupy a pair of neighboring
sites.
   Each brick has width two and the height $1$ and they 
accumulate at sites without overlap. 
  The dynamics of the brick model is described using   
the brick height $\mathcal{K}_{j}(n)$ at the $j$-th site  
after $n$ bricks have been dropped.  
   The total brick height $\mathcal{K}_{j}(n)$ takes 
on integer values 
and its dynamics is expressed as follows.
   If the $n$-th brick is dropped on sites $j_{n}$ 
and $j_{n}+1$ then
\begin{eqnarray}
   \mathcal{K}_{l}(n) 
   = \left\{
      \begin{array}{l}
         \mbox{max}\left\{\mathcal{K}_{j_{n}}(n-1), 
         \mathcal{K}_{j_{n}+1}(n-1) \right\}+ 1 \\ 
         \ls \mbox{if}\;\; l \in \{ j_{n}, 
         j_{n} + 1\} \\ \\
         \mathcal{K}_{l}(n-1)
         \;\;\; \mbox{if}\;\; 
         l \in\!\!\!\!\!/ \{ j_{n}, 
         j_{n} + 1 \}
      \end{array}
   \right.
\label{ClockValue4}\end{eqnarray}
(noting that 
the particle index $N+1$ is equivalent to $1$). 
   Here the site number $j_{n}$ is chosen randomly 
on $[1,N]$ for each $n$. 

   One may notice that the brick accumulation model 
is a one-dimensional version of the so-called 
\emph{clock model} \cite{Zon98,Bei00,Zon02}, 
which has been used to calculate the 
Lyapunov exponents for many-hard-disk systems. 
   In the clock model, the 
brick height $\mathcal{K}_{j}(n)$ is usually called 
the {\it clock value} of the $j$-th particle 
after the $n$-th particle-particle collision, 
and the dynamics (\ref{ClockValue4}) 
is expressed as an adjustment of the clock values 
of colliding particles. 
   However we use the name {\it brick accumulation model} 
for the one-dimensional version of the clock model, 
because in a simple image of brick accumulations, it is 
easily to visualize 
the configuration of brick heights which is essential 
for the analytical approach to the hopping rate discussed 
in the next section \ref{AnalyticalExpression}, 
although this image is applicable only for the 
one-dimensional case. 
   The image of brick accumulations also helps 
us to easily recognize the similarity of this model 
to ballistic aggregation models, 
whose scaling properties have been studied 
analytically and numerically \cite{Fra00,Rom02}. 

   In the brick model described by the brick height 
$\mathcal{K}_{j}(n)$, 
the site number $j$ corresponds to the particle index, 
and the number $n$ of dropped bricks corresponds 
to the collision number $n_{t}$ for the 
quasi-one-dimensional hard-disk system.  
   Moreover, the brick height $\mathcal{K}_{j}(n)$ 
itself is connected to 
the Lyapunov vector component amplitude 
$|\delta\bfGamma_{j}^{(1)}(t)|$ by 
\begin{eqnarray}
   \left|\delta\bfGamma_{j}^{(1)}(\tau n_{t})\right| \sim
   \left|\delta\bfGamma_{j}^{(1)}(0)\right|
   \exp\left\{-\mathcal{K}_{j}(n_{t})\ln \rho\right\}
\label{brikheight}\end{eqnarray}
for the particle index $j$ and the collision number $n_{t}$ 
with the mean free time $\tau$ in the asymptotic limit  
of low density $\rho\rightarrow 0$. 
   The derivation of the relation (\ref{brikheight}) 
from microscopic dynamics for hard-particle systems  
is given in Appendix 
\ref{ClockModelManyHardParticleSystems}. 

   From the relation (\ref{brikheight}) between  
the brick height and the Lyapunov vector, 
the amplitude of the localized 
Lyapunov vector components with the highest brick height 
have much larger values than the other components 
because of the huge factor $-\ln\rho$ 
appearing in the exponent of Eq. (\ref{brikheight}) 
at low density $\rho<\!<1$.  
   Therefore, to a good approximation,
the normalized amplitude $\gamma_{j}^{(n)}$ 
given by Eq. (\ref{gamma}) 
has non-zero components only for
particles corresponding to these largest amplitudes, 
as shown in Fig. \ref{figA2loczAmpDyn}.  
   The combination of Eqs. (\ref{ClockValue4}) 
and (\ref{brikheight}) also explains why the 
amplitudes of the localized components 
of the Lyapunov vectors for nearest-neighbor particles 
take almost the same value, thus appearing as a flat top 
in the localized regions in Fig. \ref{figA2loczAmpDyn}.    
   After all, the localized region of the 
Lyapunov vector represented in Eq. (\ref{brikheight}) 
is given as the site indices whose brick height 
$\mathcal{K}_{j}(n_{t})$
take the maximum value after the $n$-th brick is dropped,  
and such sites with the highest brick height 
can be calculated using the brick model 
dynamics (\ref{ClockValue4}) only, without referring 
to the hard-disk dynamics. 
   Based on these features 
we can calculate the hopping distance $h$ 
for the brick accumulation model 
(see the end of 
Appendix \ref{HoppingRateLocalizedLyapunovVector} 
for more detail of the hopping distance for the brick model), 
and therefore the hopping rate $P_{N} (h)$, similarly to the 
quasi-one-dimensional hard-disk system.

\begin{figure}[t!]
\vspfigA
\caption{
      Normalized hopping rates $P_{N} (h)/P_{N} (1)$ for 
   the brick model (circles) and 
   the quasi-one-dimensional system with $d=10^{5}$ (squares)
   as a function of the absolute value 
   of hopping distance $|h|$ for $N=50$
   as log-log plots.
      The error bars are given by  $|P_{N} (h)-P_{N} (-h)|/P_{N} (1)$.  
      The broken line is 
   the analytical expression for the hopping rate 
   discussed in Sec. \ref{AnalyticalExpression}.
   }
\label{figC2hopRateBrick}
\vspfigB\end{figure}  
%
   Figure \ref{figC2hopRateBrick} 
shows the normalized hopping rates $P_{N} (h)/P_{N} (1)$ for the brick model 
with $50$ sites (circles)
and for a quasi-one-dimensional hard-disk system 
consisting of $50$ hard-disks and $d=10^{5}$ (squares). 
   Agreement between the brick model 
and the quasi-one-dimensional hard-disk system 
for the normalized hopping rate $P_{N} (h)/P_{N} (1)$ for $|h|\leq 1$ 
is satisfactory. 
   Numerical simulations of quasi-one-dimensional hard-disk systems 
show that the normalized hopping rate $P_{N} (h)/P_{N} (1)$ 
increases very slightly as the density decreases 
as shown in Appendix \ref{HoppingRateLocalizedLyapunovVector}, so 
the deviation of $P_{N} (h)/P_{N} (1)$ in tails in the brick model 
and the hard-disk system in Fig. \ref{figC2hopRateBrick} 
may be a finite density effect.

\begin{table}[t!] 
\caption{
      Particle number (N) dependences of the ratio 
   $P_{N} (0)/P_{N} (1)$ 
   of the hopping rates at the hopping distances $h=0$ and $1$  
   in the quasi-one-dimensional hard-disk systems with $d=10^{3}$ 
   and in the brick accumulation model. }
\label{NdepenPtau}
\vspfigA
\begin{tabular}{c|cc}
\hline
\hline
    & Hard-disk model & Brick model \\
   \makebox[5em]{$N$} & 
   \makebox[9em]{$P_{N} (0)/P_{N} (1)$} & 
   \makebox[8em]{$P_{N} (0)/P_{N} (1)$}  \\
   \hline
   25 & 13.5 & 17.1 \\
   50 & 26.8 & 37.9 \\
   75 & 39.6 & 58.8 \\
   100& 52.1 & 79.7 \\
   \hline \hline
\end{tabular}
\vspfigB
\end{table}%
%
   So far we have discussed the hopping rate $P_{N} (h)$ for 
non-zero hopping distances $h\neq 0$. 
   Now we discuss the hopping rate $P_{N} (0)$ 
for zero hopping distance $h=0$. 
   Table \ref{NdepenPtau} shows the ratio $P_{N} (0)/P_{N} (1)$ 
for the hopping rates at $h=0$ and $1$ in the 
quasi-one-dimensional hard-disk system with $d=10^{3}$ 
and the brick accumulation model 
for $N=25,50,75$ and $100$.
   As shown in this table,  
the ratio $P_{N} (0)/P_{N} (1)$ depends on 
the number of particles $N$, and roughly speaking it is 
proportional to $N$. 
   To explain this linear dependence of  $P_{N} (0)/P_{N} (1)$ 
with respect to $N$, we note that in the brick model 
the hopping distance $1$ occurs only when a new brick 
is dropped at one (and not both) 
of the sites with the highest brick height, 
meaning that the hopping rate $P_{N} (1)$ should be 
approximately inversely proportional to $N$ . 
   On the other hand, the hopping distance $0$ occurs 
when a new brick is dropped at a site which 
does not have the highest nor the second highest brick height, 
or is dropped at both the sites which are currently 
the site of the highest brick height,   
meaning that the hopping rate $P_{N} (0)$ should be, roughly 
speaking, independent of $N$. 
   These considerations for $P_{N} (0)$ and $P_{N} (1)$ 
explain why the ratio $P_{N} (0)/P_{N} (1)$ is proportional to $N$. 
   However we need a more detailed consideration 
to explain the difference between the coefficient $\kappa$ 
in the relation $P_{N} (0)/P_{N} (1)\approx \kappa N$ between the 
quasi-one-dimensional hard-disk system and the brick 
accumulation model.   

\begin{figure}[t!] 
\vspfigA
\caption{
      The sum $\sum_{j=1}^{N} \mathcal{K}_{j}(n)$ 
   of brick heights $\mathcal{K}_{j}(n)$ as a function of 
   the number $n$ of dropped bricks 
   in the brick accumulation model with $N=50$ 
   in the $n$ interval $[0,20979999]$.  
      Inset: The same graph but enlarged in 
   the much smaller $n$ interval $[1000,1050]$. 
      The broken line (which is almost on the data points 
   of the main graph) 
   is a fit of numerical data to  
   a linear function. 
   }
\label{figC3brickSum}
\vspfigB\end{figure}  
%
   Before finishing this section, we discuss one more property 
of the brick accumulation model, 
which we will use in the next section. 
   Fig. \ref{figC3brickSum} 
is the sum $\sum_{j=1}^{N} 
\mathcal{K}_{j}(n)$ of brick heights as a function of $n$. 
   Note that this sum must increase at a speed 
of more than 2 per dropped brick, 
because accumulated bricks capture spaces 
below them which cannot be occupied by further 
dropped bricks.   
   It is clear from this figure that this sum 
increases linearly, and a fit of the data to the function 
$\sum_{j=1}^{N} \mathcal{K}_{j}(n) =\alpha + \beta n$ 
with fitting parameters $\alpha$ and $\beta$
leads to the values $\alpha\approx 0.00$ and 
$\beta\approx 3.99$ \cite{noteSum}. 
   (Note that data points in the main graph 
of Fig. \ref{figC3brickSum} look exactly like 
this fit because of the large scale.  
   In the inset to Fig. \ref{figC3brickSum} 
we show the graph of $\sum_{j=1}^{N} \mathcal{K}_{j}(n)$ 
as a function of $n$ on a much smaller scale, 
to show its fluctuating behavior.)
   Therefore, on average, each dropped brick  
adds a contribution of 4 to the sum $\sum_{j=1}^{N} 
\mathcal{K}_{j}(n)$, meaning that each brick occupies 
not only its own $2$ spaces but also captures $2$ empty 
spaces below it. 

   It is important to note that from Eq. (\ref{brikheight}) 
the brick height 
$\mathcal{K}_{j}(n)$ dominates the exponential growth rate 
of the Lyapunov vector component amplitude.  
   Using this feature of the brick heights, 
their sum $\sum_{j=1}^{N} \mathcal{K}_{j}(n)$ 
is connected to the largest Lyapunov exponent $\lambda^{(1)}$ as 
\begin{eqnarray}
   \lambda^{(1)} \sim -\lim_{n\rightarrow +\infty} 
   \frac{1}{n\tau} \left[\frac{1}{N}\sum_{j=1}^{N}
   \mathcal{K}_{j}(n) \right]\ln\rho
\label{largeLyapu}\end{eqnarray}
with the mean free time $\tau$ and the density $\rho$ 
in the asymptotic limit of low density.  
   [More detailed discussion for the formula (\ref{largeLyapu}) 
is given in Appendix \ref{ClockModelManyHardParticleSystems}.]
   As a numerical check of the formula (\ref{largeLyapu}) 
we show in Fig. \ref{figC4LyapuMax}, 
the largest Lyapunov exponent $\lambda^{(1)}$ as 
a function of the density $\rho$ in a quasi-one-dimensional 
hard-disk systems and in the brick accumulation model 
using the formula (\ref{largeLyapu}) 
with $N=50$. 
   Here, to calculate the largest Lyapunov exponent
from Eq. (\ref{largeLyapu}) 
we used the relation $\sum_{j=1}^{N}
\mathcal{K}_{j}(n)\approx 4n$ as shown 
in Fig. \ref{figC3brickSum}, 
and  the values of the mean free time $\tau$ and 
the density $\rho$ of the quasi-one-dimensional 
system whose Lyapunov exponents are plotted 
in Fig. \ref{figC4LyapuMax}, and 
the data points are connected by a dashed line for ease of visibility. 
   Figure \ref{figC4LyapuMax} shows that 
Eq. (\ref{largeLyapu}) reproduces 
successfully the values of the largest Lyapunov exponent 
for the quasi-one-dimensional hard-disk systems, 
not only in the limit of low density 
but also at relatively high density such as $\rho < 0.3$. 
   It may be noted that a linear dependence of 
the sum $\sum_{j=1}^{N} \mathcal{K}_{j}(n)$ of brick heights 
with respect to $n$  
is necessary to get a finite value of the largest Lyapunov 
exponent $\lambda^{(1)}$ by Eq. (\ref{largeLyapu}). 
%
\begin{figure}[t!] 
\vspfigA
\caption{
      The largest Lyapunov exponent $\lambda^{(1)}$ as 
   a function of density $\rho$ in a quasi-one-dimensional 
   hard-disk system with $N=50$ as a log-log plot. 
      The error bars are given by $|\lambda^{(1)}-\lambda^{(4N)}|$ 
   which must be zero by the conjugate pairing rule 
   for Hamiltonian systems. 
      The broken line is the largest Lyapunov exponent 
   given by the brick accumulation model with $N=50$.   
   }
\label{figC4LyapuMax}
\vspfigB\end{figure}  
%

\section{Analytical Expression for the Hopping Rate}
\label{AnalyticalExpression}

   In this section we discuss another approach 
to the hopping rate of the localized region of the Lyapunov vectors. 
   This approach is inspired by some of the characteristics 
of the brick accumulation model discussed in Sec. \ref{brickmodel}, 
although we greatly simplified the brick model 
by omitting some other aspects of the model.  
   The advantage of this approach is that we can get an analytical 
expression for the hopping rate, while to a good approximation 
it still reproduces the hopping rate for hard-disk systems.  
   Another important point in this approach 
is that it connects the dynamics of the localized region of 
the Lyapunov vectors with a static property,  
the probability distribution of brick-height differences 
between nearest-neighbor sites.

   In the brick accumulation model, 
a hop of the highest brick site 
occurs when two (non-nearest-neighbor) sites have 
the same brick height. 
   Noting this characteristic we consider the 
probability distribution $\tilde{P}_{\infty}(h)$ 
under the constraint that the two sites $\mu$ and $\mu+h$ have the 
highest brick height. 
   We require that there is no other highest site 
between the two highest sites $\mu$ and $\mu+h$ for $|h|\geq 2$;  
\begin{eqnarray}
   && \hspace{-0.5cm}
   \mathcal{K}_{l}(n) < \mathcal{K}_{\mu}(n) 
      \;\;\;[=\mathcal{K}_{\mu+h}(n)],  \nonumber \\
   &&\ls
   \mbox{in} \;\;\; l=\left\{\begin{array}{l}
      \mu+1,\mu+2,\cdots,\mu+h-1   \\
      \ls \mbox{for} \;\;\; h\geq 2 \\
      \mu+h+1,\mu+h+3,\cdots,\mu-1 \\
      \ls  \mbox{for} \;\;\; h\leq -2,  
   \end{array}\right.
\label{CondiPassInte}\end{eqnarray}
   Using this probability distribution 
$\tilde{P}_{\infty}(h)$, we can estimate 
the hopping rate $P_{\infty}(h)$ 
in the thermodynamic limit as the one proportional to 
$\tilde{P}_{\infty}(h)$:  
$P_{\infty}(h) 
\propto \tilde{P}_{\infty}(h)$.  
   Then we can calculate the hopping rate $P_{N} (h)$ 
for a finite size system using Eq. (\ref{normalizedP}), 
apart from a constant factor.

\begin{figure}[t!]
\vspfigA
\caption{
   Schematic illustration for a brick-height 
   configuration (similar to Fig. \ref{figC1brickModel} 
   for the brick model)
   connecting two highest sites 
   (particle indices) $\mu$ and $\mu+h$ with 
   hopping distance $h$ in the case of $h\geq 2$.  
      Here, $-k_{l}$ is the brick-height difference 
   of the sites $\mu+l-1$ and $\mu+l$ 
   ($l=1,2,\cdots,h-1$).  
      The probability distribution for 
   the configuration with $k_{l}$, $l=1,2,\cdots,h-1$
   is given by 
   $\Lambda(-k_{1})  \Lambda(-k_{2}) \Lambda(-k_{3})
   \cdots \Lambda(-k_{h-1}) \Lambda(\sum_{j=1}^{h-1} k_{j})$ 
   with the distribution $\Lambda(k)$ 
   of brick-height differences $k$ between 
   nearest-neighbor sites. 
     The hopping rate $\tilde{P}_{\infty}(h)$ 
  is calculated as the summation of this probability distribution 
  over possible values of $k_{l}$, $l=1,2,\cdots,h-1$.
   }
\label{figD1brickConfig}
\vspfigB\end{figure}  
%
   To calculate the probability distribution 
$\tilde{P}_{\infty}(h)$ 
we introduce the distribution $\Lambda(k)$ 
of brick-height differences $k$ between 
nearest-neighbor sites. 
   These two distribution functions are connected by   
%
\begin{eqnarray}
   \tilde{P}_{\infty}(h) &=& 
      \sum\limits_{
            \sumindex{k_{1}}{k_{1}\geq 1}}
         \sum\limits_{
            \sumindex{k_{2}}{k_{1}+k_{2}\geq 1}}  
         \cdots
         \sum\limits_{
            \sumindex{k_{|h|-1}}{\sum_{j=1}^{|h|-1} k_{j}\geq 1}}
       \nonumber \\
       &&\hspace{0cm} \times
      \Lambda(-k_{1})  \Lambda(-k_{2}) 
       \cdots 
            \Lambda(-k_{|h|-1}) 
            \Lambda\left(\sum\limits_{j=1}^{|h|-1} k_{j}\right) .
       \nonumber \\
\label{passInte1}\end{eqnarray}
%
   In Eq. (\ref{passInte1}), $-k_{l}$ is a brick-height difference 
of nearest neighbor sites 
($l=1,2,\cdots,|h|-1$).  
      The probability distribution for 
the specific brick-height configuration 
with $k_{l}$, $l=1,2,\cdots,|h|-1$ is given by 
$\Lambda(-k_{1})  \Lambda(-k_{2}) \Lambda(-k_{3})
\cdots \Lambda(-k_{|h|-1}) \Lambda(\sum_{j=1}^{|h|-1} k_{j})$, 
and the hopping rate $\tilde{P}_{\infty}(h)$ 
is calculated as the summation of this probability distribution 
over possible values of $k_{l}$, $l=1,2,\cdots,h-1$, 
leading to Eq. (\ref{passInte1}).   
   Figure \ref{figD1brickConfig} is 
a schematic illustration of a brick-height 
configuration connecting two highest sites and 
the brick-height differences $-k_{j}$ of 
nearest neighbor sites 
in the case of $h\geq 2$. 
   In Eq. (\ref{passInte1}), 
the lower bounds of the sums on the right-hand 
side of Eq. (\ref{passInte1}) 
come from the condition (\ref{CondiPassInte}). 
   Note that from Eq. (\ref{passInte1}) 
the distribution $\tilde{P}_{\infty}(h)$ 
is an even function of $h$, namely $\tilde{P}_{\infty}(-h) 
= \tilde{P}_{\infty}(h)$.

   Now, for simplicity, we assume that the distribution function 
$\Lambda(k)$ can be expressed as an exponential function 
\begin{eqnarray}
   \Lambda(k) = \mathcal{W}\exp\{-\eta |k|\}
\label{BrickDisNear}\end{eqnarray}
with constants $\mathcal{W}$ and $\eta \; (>0)$. 
   (We will discuss the validity of this assumption later.)
   Inserting Eq. (\ref{BrickDisNear}) into 
Eq. (\ref{passInte1}), and replacing the sums over 
$k_{1}$ 
in Eq. (\ref{passInte1}) with the ones over 
$k_{1}'\equiv k_{1}+1$, we obtain 
%
%
\begin{eqnarray}
   \tilde{P}_{\infty}(h) 
   = \left\{\begin{array}{l} 
      \mathcal{W}^{2} \Upsilon
      \sum\limits_{k=0}^{+\infty} \Upsilon^{k}
          \hspace{2.6cm} \mbox{for}\;\; |h|=2 \\ \\
      \mathcal{W}^{|h|}\Upsilon
      \sum\limits_{k_{1}=0}^{+\infty} \; \Upsilon^{k_{1}} 
          \!\!\!\sum\limits_{k_{2}=-k_{1}}^{+\infty}  
             \!\!\!\! \Upsilon^{k_{2}\theta(k_{2})} 
             \!\!\!\!
             \sum\limits_{k_{3}=-k_{1}-k_{2}}^{+\infty} 
             \!\!\!\!\!\!\!\! 
             \Upsilon^{k_{3}\theta(k_{3})}  \\
         \ls\ls \cdots \!\!\!\!
            \sum\limits_{k_{|h|-1}
               =-\sum\limits_{j=1}^{|h|-2}k_{j}}^{+\infty}     
               \!\!\!\!\!\!
               \Upsilon^{k_{|h|-1}\theta(k_{|h|-1})} \\
        \hspace{4.5cm} \mbox{for}\;\; |h|\geq 3
   \end{array}\right. \nonumber
\end{eqnarray} \vspace{-0.5cm}
\begin{eqnarray}
\label{passInte2}\end{eqnarray}
where $\Upsilon$ is defined by 
\begin{eqnarray}
   \Upsilon\equiv \exp\{-2\eta\}
\label{Upsilon}\end{eqnarray}
and $\theta(x)$ is the Heaviside function taking the value $1$ 
for $x>0$ and the value $0$ for $x\leq0$. 
   The summations appearing in Eq. (\ref{passInte2}) can be 
carried out successively, and we obtain 
\begin{eqnarray}
   \tilde{P}_{\infty}(h) 
   = \frac{\mathcal{W}^{|h|}\Upsilon}{(1-\Upsilon)^{|h|-1}} \Omega(|h|)
\label{passInte3a}\end{eqnarray}
where the function $\Omega(k)$ of $k$ is can be written as  
\begin{eqnarray}
     \Omega(2) &=& 1 , 
        \label{omega2} \\
     \Omega(3) &=& 1 + \Upsilon , 
        \label{omega3} \\
     \Omega(4) &=& 1 + 3\Upsilon + \Upsilon^2 ,
        \label{omega4} \\
     \Omega(5) &=& (1 + \Upsilon)(1 + 5\Upsilon + \Upsilon^2) ,
        \label{omega5} \\
     \Omega(6) &=& 1 + 10\Upsilon + 20\Upsilon^2 
        + 10\Upsilon^3 + \Upsilon^4  ,
        \label{omega6}\\
     \Omega(7) &=& (1 + \Upsilon)(1 + 14\Upsilon 
        + 36\Upsilon^2 + 14\Upsilon^3 + \Upsilon^4) , \ls  
        \label{omega7} 
\label{passInte3b}\end{eqnarray}
and so on. 
   Eq. (\ref{passInte3a}), with Eqs. 
(\ref{omega2}), (\ref{omega3}), (\ref{omega4}), (\ref{omega5}), 
(\ref{omega6}), (\ref{omega7}), etc., 
gives an analytical expression 
for the hopping rate, for example, using the relation 
$P_{\infty}(h) /P_{\infty}(1) =
\tilde{P}_{\infty}(h) /\tilde{P}_{\infty}(1)$.  

   The coefficients $\mathcal{W}$ and $\eta$ 
appearing in Eq. (\ref{BrickDisNear}) 
can be determined from the two sum rules:  
\begin{eqnarray}
   \sum_{k=-\infty}^{+\infty}\Lambda(k) &=& 1 , 
      \label{sumRole1}\\
   \sum_{k=-\infty}^{+\infty}|k| \Lambda(k)&=&\Delta\mathcal{K},  
      \label{sumRole2}
\end{eqnarray}
where $\Delta\mathcal{K}$ is the mean value of 
the absolute value of the 
brick height difference between nearest-neighbor sites. 
   The first condition (\ref{sumRole1}) is  
the normalization of the probability distribution 
$\Lambda(k)$ which leads to 
\begin{eqnarray}
   \mathcal{W} 
   = \frac{\exp\{\eta\}-1}{\exp\{\eta\}+1} .
\label{normLambda1}\end{eqnarray}
   Using Eq. (\ref{normLambda1}) 
the second condition (\ref{sumRole2}) gives  
\begin{eqnarray}
   \eta = \ln\left\{
      \frac{
      1 +\sqrt{1+\Delta\mathcal{K}^{2}}
      }{\Delta\mathcal{K}} \right\} ,
\label{normLambda2}\end{eqnarray}
satisfying the inequality $\eta>0$. 
   Inserting Eq. (\ref{normLambda2}) into Eqs. 
(\ref{Upsilon}) and (\ref{normLambda1}) we obtain 
\begin{eqnarray}
   \Upsilon &=& \left(
      \frac{\Delta\mathcal{K}}
      {1 +\sqrt{1+\Delta\mathcal{K}^{2}}}
      \right)^{2}
      \label{UpsilonB} \\
   \mathcal{W} &=&  
      \frac{1 +\sqrt{1+\Delta\mathcal{K}^{2}}
         -\Delta\mathcal{K}}
      {1 +\sqrt{1+\Delta\mathcal{K}^{2}}
         +\Delta\mathcal{K}} .
\label{normLambda1b}\end{eqnarray}
   From Eqs. (\ref{UpsilonB}) and (\ref{normLambda1b}),   
there is only one parameter $\Delta\mathcal{K}$ 
remaining to determine 
the hopping rate using Eq. (\ref{passInte3a}).

   To estimate the value of $\Delta\mathcal{K}$, 
we use a property of the brick accumulation model 
discussed at the end of Sec. \ref{brickmodel}. 
   Previously,  
we showed an approximate relation 
$\sum_{j=1}^{N}\mathcal{K}_{j}(n)\approx 4n$ 
for the sum of brick heights, 
which means that in the brick accumulation model 
each dropped brick gives a mean  
contribution of 4 to this sum. 
   This contribution consists of 
a contribution of 2 as the space occupied  
by a brick itself 
and  another 2 as empty space below the brick
which can now not be occupied by other
bricks. 
   This implies that the averaged 
brick-height difference between nearest-neighbor sites 
is about 2, so that  
\begin{eqnarray}
   \Delta\mathcal{K} \approx 2. 
\label{paramDK}\end{eqnarray}
   We use this value to calculate the hopping rate 
based on Eq. (\ref{passInte3a}). 
   One may notice that from the relation 
$\sum_{j=1}^{N}\mathcal{K}_{j}(n) 
\approx (\Delta\mathcal{K}+2)n$ and 
the formula (\ref{largeLyapu}) we obtain 
\begin{eqnarray}
   \Delta\mathcal{K} \sim 
   \frac{N\tau}{\ln\rho} \lambda^{(1)}-2 
\label{largeLyapu2}\end{eqnarray}
in the limit of low density. 
   From the relation (\ref{largeLyapu2}) 
between the parameter $\Delta\mathcal{K}$ specifying 
the hopping rate and the largest Lyapunov exponent 
$\lambda^{(1)}$, the hopping rate of the localized region 
of the Lyapunov vectors is connected to the 
largest Lyapunov exponent. 

   Before comparing the hopping rate  
based on the analytical expression  (\ref{passInte3a}) 
with the ones for hard-disk systems and the brick 
accumulation model, we note that this analytical 
expression for the hopping rate may not be appropriate 
for small hopping distances $|h|$. 
   In particular, it does not give the correct value of 
$P_{N} (1)$, because in the brick model 
the hopping distance $h=\pm 1$ does not appear from 
separated highest sites with the same brick height, 
the assumption used to derive Eq. (\ref{passInte3a}). 
   Therefore it is not appropriate to calculate 
the hopping rate normalized by $P_{N} (1)$ from 
this analytical expression and 
to compare it with the numerical results. 
   The hopping rate $P_{N} (\pm 2)$ from 
Eq. (\ref{passInte3a}) may also be problematic, 
because in the brick model the hopping distance 
$h=\pm 2$ occurs when 4 consecutive sites have 
the highest brick height, while to derive 
Eq. (\ref{passInte3a}) we assumed that 
$h=\pm 2$ occurs when non-consecutive 
separate sites $\mu$ and $\mu+h$ 
have the same highest brick height. 
   Based on these considerations we do not calculate  
the value $P_{N} (\pm 1)$ from the analytical approach 
in this section, 
and plot the hopping rate so that $P_{N} (5)$ 
by Eq. (\ref{passInte3a}) 
gives the same value as that from the brick model. 

   In Fig. \ref{figC2hopRateBrick} 
for the normalized hopping rate $P_{N} (h)/P_{N} (1)$, 
we plotted the function 
$\Psi(h) \equiv \tilde{P}_{N}(h)P_{N} (5)
/[\tilde{P}_{N}(5)P_{N} (1)]$  
using the value $P_{N} (5)/P_{N} (1)$ of the brick model with 
   $\tilde{P}_{N}(h) = \tilde{P}_{\infty}(h)
   +\tilde{P}_{\infty}(N-h)$ 
   using the analytical expression 
   $\tilde{P}_{\infty}(h)$  
given by Eq. (\ref{passInte3a})  for $|h|\geq 2$. 
   Note $\Psi(5)= P_{N} (5)/P_{N} (1)$ so that $\Psi(h)$ 
coincides with $P_{N} (h)/P_{N} (1)$ of the brick model at $|h|=5$. 
   Here, the value of the hopping rate values 
are given at integer values of $h$, 
but we connect them with a broken line for ease of visibility. 
   Figure \ref{figC2hopRateBrick} shows that 
the analytical expression (\ref{passInte3a}) 
for the hopping rate reproduces the hopping rate 
for the brick accumulation model as well as 
the quasi-one-dimensional hard-disk system 
to a good approximation. 
   It may be noted that for this plot we used 
the first two dominant terms on the right-hand side of 
Eq. (\ref{normalizedP}), and part of the small deviation of 
the hopping rate in the tail between the brick model 
and the analytical expression should come from the 
omission of higher order terms in Eq. (\ref{normalizedP}).

\begin{figure}[t!]
\vspfigA
\caption{
   The distribution $\Lambda(k)$ of the brick-height differences $k$ 
   between nearest-neighbor sites as a function of $|k|$ 
   in the brick accumulation 
   model with $N=50$ as a linear-log plot. 
      Here,  $\Lambda(k)$ is normalized 
   by $\sum_{k=-2N}^{2N}\Lambda(k)=1$.
      The distribution $\Lambda(k)$ is an even function of $k$, 
   and the error bars in this graph 
   are given by $|\Lambda(k)-\Lambda(-k)|$. 
      The line is an exponential function 
   used to obtain the analytical expression for the hopping rate 
   given by Eqs. (\ref{BrickDisNear}), (\ref{UpsilonB}),
   (\ref{normLambda1b}) and (\ref{paramDK}). 
   }
\label{figD2brickDisNear}
\vspfigB\end{figure}  
%
   We notice that the approach in this section is simple 
enough to get an analytical expression for the hopping rate 
but it is not completely consistent 
with the brick accumulation model 
discussed in the previous section \ref{brickmodel}.  
   Previously, we have mentioned 
the irrelevance of Eq. (\ref{passInte3a}) as a description 
of a small hopping rate in the brick model.
   Actually, in the approach of this section we omitted 
the essential characteristic of the bricks 
as components of the brick 
accumulation model, 
except for the property (\ref{paramDK}), 
and treated 
the model components as blocks (or half bricks). 
   As another example, we show in Fig. \ref{figD2brickDisNear} 
the numerical result 
for the distribution $\Lambda(k)$ of brick-height differences $k$ 
between nearest-neighbor sites in the brick accumulation 
model with $N=50$.
   Here, $\Lambda(k)$ is normalized 
by $\sum_{k=-2N}^{2N}\Lambda(k)=1$, instead of Eq. (\ref{sumRole1}), 
because we cannot calculate $\Lambda(k)$  
in $|k|\rightarrow +\infty$ numerically.
   In this figure we added the exponential distribution  (\ref{BrickDisNear}) using Eqs. (\ref{UpsilonB}),
   (\ref{normLambda1b}) and (\ref{paramDK}). 
   Figure \ref{figD2brickDisNear} shows that the 
probability distribution $\Lambda(k)$ does not coincide 
with an exponential distribution (\ref{BrickDisNear}), 
although it may be justified as a first approximation. 
   (On the other hand, the numerical evaluation of 
$\Delta\mathcal{K}$ from the distribution 
$\Lambda(k)$ as a numerical result 
in Fig. \ref{figD2brickDisNear} gives the value 
$1.99$, then Eq. (\ref{paramDK}) is still justified.)
   On another point, in the brick model the highest brick sites 
appear as a pair of nearest-neighbor sites, but we did not 
take into account this characteristic in the analytical 
approach in this section. 
   Despite the omission of characteristics of the 
brick accumulation model, the analytical approach in this section 
reproduces the hopping rate for many-hard-disk systems 
reasonably well, and it suggests 
that this approach still keeps enough 
of the essential characteristics 
that describe the dynamics of the Lyapunov localization.

\section{Conclusion and Remarks}
\label{ConclusionRemarks}

   In this paper we discussed the dynamics of the spatially 
localized region of the Lyapunov vector corresponding 
to the largest Lyapunov exponent in a quasi-one-dimensional 
hard-disk system. 
   To discuss the dynamics of the localized region 
of the Lyapunov vector 
we introduced a hopping rate for the localized region, 
and showed that the hopping rate decreases as the absolute value 
of hopping distance increases. 
   This hopping-distance dependence of the hopping rate 
was explained quantitatively in two ways: 
a brick accumulation model and an analytical approach. 
   In the brick accumulation model, the hopping behavior 
of the localized Lyapunov vectors was explained as 
the movement of the highest position in the 
brick accumulations. 
   It was shown that using this brick model 
we can calculate the largest Lyapunov exponent 
for quasi-one-dimensional hard-disk systems 
successfully. 
   On the other hand, in the analytical approach 
the hopping rate was calculated from probability 
distributions for brick height differences of 
nearest neighbor sites via multiple summations  
over possible configurations that connect two separated 
highest sites. 
   The result is related to the largest Lyapunov exponent.
   Both of the approaches successfully reproduced 
the hopping-distance dependence of the hopping rate 
for the localized Lyapunov vectors of 
quasi-one-dimensional hard-disk systems.

\begin{figure}[t!]
\vspfigA
\caption{
      The distribution $D(\bfsigma\cdot\Delta\bfp)$ 
   of the quantity 
   $\bfsigma\cdot\Delta\bfp$ with 
   the normalized collision vector $\bfsigma$ and 
   the momentum difference $\Delta\bfp$ of colliding particles 
   just before the collision; 
   for the general case (dashed line) 
   and for the case in which non-zero hopping 
   of a localized Lyapunov vector occurs. 
      The system is a quasi-one-dimensional system 
   with $d=10^{3}$ and $N=25$. 
   }
\label{figE1ndpdistri}
\vspfigB\end{figure}  
%
   As a remark, it may be useful to mention 
a previous conjecture for the dynamics of the localized region 
of Lyapunov vectors corresponding to largest Lyapunov exponent. 
   Before we started this work, 
there had been a view that the origin of the hopping 
behavior of localized Lyapunov vectors was: 
\begin{description}
\item[$\langle${\it Conjecture}$\rangle$] 
   {\it The localized region of the Lyapunov vector hops 
   to the position
   of a new grazing collision.}
\end{description}
   This conjecture was suggested from the fact that a change 
of Lyapunov vectors in particle collisions may be  
dominated by the factor $1/(\bfsigma\cdot\Delta\bfp)$ in the 
collision dynamics for Lyapunov vector (see Appendix 
\ref{ClockModelManyHardParticleSystems}), where $\bfsigma$ is   
the normalized collision vector and $\Delta\bfp$ is  
the momentum difference of the colliding particles before 
the collision. 
   In this argument, if two particles collide at a small angle 
(a grazing collision) with a small value of   
$\bfsigma\cdot\Delta\bfp$, then the Lyapunov vector can change  
significantly and the position of 
the localized region may move. 
   However, this scenario cannot be correct
for the following reasons. 
   First, this conjecture implies that as the position at 
which a grazing collision occurs is random,  
the hopping distance should also be random.  
   This contradicts our numerical result that  
the hopping rate decreases with increasing hopping distance,
as shown in Fig. \ref{figB1hopRateN}.  
   Second, we show in
Fig. \ref{figE1ndpdistri}, the distribution  
$D(\bfsigma\cdot\Delta\bfp)$ of $\bfsigma\cdot\Delta\bfp$ 
in general collisions (dashed line),
and the distribution of collisions causing hopping of the 
localized Lyapunov vector (solid line) \cite{noteB}. 
   It is clear from Fig. \ref{figE1ndpdistri} that 
the quantity $\bfsigma\cdot\Delta\bfp$ is not smaller 
in collisions which lead to jumps of the localized region of 
Lyapunov vector compared with the general case.  
   It may also be noted that the hopping dynamics 
of the Lyapunov vector from the brick accumulation model 
described in Sec. \ref{brickmodel} is 
independent of collision parameters like collision angles,  
the momentum difference of colliding particles, and also  
the quantity $\bfsigma\cdot\Delta\bfp$. 
   This also suggests that the conjecture  
for the origin of the hopping of the localized region of 
the Lyapunov vector cannot be justified.  
   On the other hand, this collision parameter independence 
of the hopping behavior in the brick model 
cannot explain why the distribution 
$D(\bfsigma\cdot\Delta\bfp)$ is different from  
the general case and the hopping case 
in Fig. \ref{figE1ndpdistri}. 
   This is an open problem. 
   
   As another open problem, the fits in Fig. \ref{figB1hopRateN} 
suggest that the hopping-distance dependence of the hopping rate 
in the thermodynamic limit seems to be a power law: 
$P_{\infty}(h)\sim h^{\beta}$ 
with $\beta\approx 1.7$. 
   It remains to be determined whether this power behavior 
of the hopping rate can be justified analytically 
in the brick accumulation model or the analytical approach 
discussed in Sec. \ref{AnalyticalExpression}. 

   In this paper we showed that the hopping rate of the localized 
region of a Lyapunov vector, and the largest Lyapunov exponent,   
are well described by the brick accumulation model. 
   Then, one may ask a more direct numerical check of 
the justification of the brick accumulation model 
in quasi-one-dimensional hard-disk systems, 
for example, to check Eq. (\ref{brikheight}) 
or to observe numerically an actual brick 
configuration like that presented in Fig. \ref{figC1brickModel}. 
   However, such a check of the brick model is not trivial 
for the following reasons. 
   First, the brick accumulation model is only justified in the 
limit of low density, but numerical simulations have to
be at some finite density.  
   This effect appears, for example, as gradual changes 
of the Lyapunov vector component amplitudes, as shown 
in Fig. \ref{figX1loczAmpDynGradu}, which do not 
appear in the brick accumulation model. 
   Second, even if we could simulate at an extremely 
low density in which such finite density 
effects can be neglected (although the case presented in 
this paper is not at such a low density), 
the factor 
$-\ln\rho$ in Eq. (\ref{brikheight}) may be too 
large for an actual numerical calculation. 
   Finally, to calculate Lyapunov vectors in this paper  
we used the algorithm developed by Benettin \textit{et al} 
\cite{Ben80a,Ben80b}. 
   This algorithm includes intermittent renormalizations 
of Lyapunov vectors, to prevent a divergence of the 
amplitudes of Lyapunov vectors in numerical calculations, 
but the brick accumulation model 
does not have such a normalization procedure in its dynamics.  
   This difference makes a direct numerical check of 
Eq. (\ref{brikheight}) difficult in hard-disk systems. 
   Different from Eq. (\ref{brikheight}) itself, 
the localized region of the Lyapunov vectors, 
which is required to calculate the hopping rate, 
is given simply by particle indices with the largest 
Lyapunov vector component amplitude, 
which is not influenced by such a difference of normalization 
procedure in calculations of Lyapunov vectors. 

   In this paper, in order to introduce the hopping rate 
we used the property of quasi-one-dimensional systems, 
that the order of particles is an invariant. 
   In this sense, it is not trivial to generalize 
our argument to fully two- (or three-) 
dimensional systems. 
   An effective way to describe 
the dynamics for the localized region of the Lyapunov vectors  
for higher spatial dimensions  
remains an important future problem. 
   Related to this  
it should still be noted that it is known that 
the clock model version of the brick accumulation model 
itself can be easily generalized to 
any spatial dimensional case, although 
in higher dimension we do not have the concept  
of the accumulation of bricks, as we do for the 
quasi-one-dimensional case.  
 
   Finally, one should notice that the brick accumulation model 
(more generally the clock model) 
used in this paper has been justified 
for hard-disk (or hard-sphere) systems only (at least so far). 
   On the other hand, the Lyapunov localization is observed 
not only in many-hard-disk systems but also 
in a wide variety of many-body chaotic systems, such as 
the Kuramoto-Sivashinsky model \cite{Man85}, 
a random matrix model \cite{Liv89}, 
map systems \cite{Kan86,Fal91,Gia91,Pik98}, 
coupled nonlinear oscillators \cite{Pik01}, etc. 
   It should be an important future problem 
to develop approaches to the dynamics of Lyapunov localization 
in this wider class of systems.




\section*{Acknowledgements}

   One of the authors (T.T) wishes to thank to R. van Zon, 
who indicated the possibility 
of describing the hopping dynamics for the localized Lyapunov vectors 
using the clock model, and supplied a computer program 
for the clock model. 
   We also appreciate helpful discussions 
with E. Zabey and J. -P. Eckmann.
   In particular, they suggested the importance of 
the difference of brick heights for 
nearest-neighbor sites in the calculation of 
the hopping rate, and posed
 a question about the behavior of the sum of brick 
heights as shown in Fig. \ref{figC3brickSum}, 
and also indicated the analogy between the brick accumulation 
model and ballistic aggregation models. 
   We also thank H. A. Posch for a comment on 
Lyapunov localization, which sparked our interest in
this problem.  

   The authors appreciate the financial 
support of the Japan Society for the Promotion of Science.

\appendix
\section{Hopping Rate of a Localized Lyapunov Vector}
\label{HoppingRateLocalizedLyapunovVector}

   In this appendix we give the detailed definition of the hopping 
rate for the localized region of the Lyapunov vectors, which is used 
in this paper.  



   As suggested in Ref. \cite{Tan03b,Tan05b}, 
the normalized Lyapunov vectors corresponding to the large 
Lyapunov exponents have 
non-zero components for only two particles  
in the low density limit, and changes of the non-zero components 
are caused by particle-particle collisions. 
   Applying this characteristic of the Lyapunov vector 
to the quasi-one-dimensional system with periodic boundary 
conditions, we can introduce a hopping distance $h^{[n]}$ 
at the $n$-th particle-particle collision as 
\begin{eqnarray}
   h^{[n]} = j_{n+1} - j_{n} - N \; 
             \underline{\mbox{nint}}\left\{\frac{j_{n+1} - j_{n}}{N}\right\} 
\label{hopDista0}\end{eqnarray}
in the low density limit. Here $\{j_{n},j_{n}+1\}$ are the non-zero
components before the $n$-th collision and 
$\{j_{n+1},j_{n+1}+1\}$ are the non-zero components 
just after the $n$-th collision. The function 
$\underline{\mbox{nint}}\{x\}$ is the closest integer 
to the real number $x$. 
   We assume that changes in the position of the
localized region of the Lyapunov 
vector are negligible during the free-flight interval, 
so $\{j_{n+1},j_{n+1}+1\}$ can also be interpreted as
the set of the particle indices whose Lyapunov vector 
components take non-zero values just before the $(n+1)$-th 
collision.
   We count the number of times $N_{T}(h)$ that we see a hop of 
size $h$  in a time-interval $T$ where 
$ -[N/2] \leq h \leq [N/2]$. 
The normalized hopping rate $P_{N} (h)$ can be introduced as 
$P_{N} (h)/P_{N} (1) = \lim_{T\rightarrow \infty}N_{T}(h)/N_{T}(1)$. 


\begin{figure}[t!]
\vspfigA
\caption{
      The normalized amplitude $\gamma_{j}^{(1)}$ of the 
   Lyapunov vector component $\delta\bfGamma_{j}^{(1)}$ 
   corresponding to the largest Lyapunov exponent $\lambda^{(1)}$
   as a function of the collision number $n_{t}$ 
   and the particle index $j$ in a quasi-one-dimensional 
   system with $d=10^{4}$ and $N=50$. 
      Here, the collision number interval shown in this graph is 
   $[k,k+68]$ with $k=3600448$. 
      On the base of this graph is a contour plot of  
   $\gamma_{j}^{(1)}$ at the level $0.2$. 
      Three hops of the localized region 
   of the Lyapunov vector are visible in this time interval; 
   The first two hops are sharp and the last one 
   is gradually.  
   }
\label{figX1loczAmpDynGradu}
\vspfigB\end{figure}  
%
   However, in actual numerical simulations, 
the particle density $\rho$ is always finite, 
and non-zero Lyapunov components of more than two particles 
can often be seen,  
at least down to a density $\rho\approx 10^{-5}$ 
which is the low density limit of the numerical simulations 
in this paper.  
   This makes the above definition (\ref{hopDista0}) 
of the hopping distance $h^{[n]}$ impractical.  
   To explain this point concretely 
we show Fig. \ref{figX1loczAmpDynGradu}, 
which is a graph of the normalized 
Lyapunov vector component amplitude $\gamma_{j}^{(1)}$ 
defined by Eq. (\ref{gamma}) 
as a function of the collision number $n_{t}$ and 
the particle index $j$ in a quasi-one-dimensional 
system with $N=50$ and $d=10^{4}$, (namely a density 
$\rho \approx 7.85\times 10^{-5}$). 
   In this figure we can recognize 
three types of hops of the localized region of the Lyapunov vector.
   The first two hops  keep the
non-zero Lyapunov vector components of almost two particles 
sharp enough 
to apply the definition (\ref{hopDista0}) of the hopping distance, 
but the third hop occurs gradually so that no 
clear hopping time can be determined. 
   Note that the localized region of 
Lyapunov vector component amplitudes in the 
three dimensional plot \ref{figX1loczAmpDynGradu} 
always has a flat top with a width of two-particles 
even for the third hop in Fig. (\ref{figX1loczAmpDynGradu}).

   In this paper, for concrete calculations, 
we define the localized region of the 
Lyapunov vector components as the particle indices $j$ 
for which $\gamma_{j}^{(n)}>0.2$. 
   (Here, we use the similarity between 
particle positions and particle indices 
given in Fig. \ref{figA1qua1dim}  
for the quasi-one-dimensional system.) 
   In Fig. \ref{figX1loczAmpDynGradu}, this localized 
region is approximately given by the region surrounded 
by the contour lines (level $0.2$) on the base of the graph. 
   We introduce the quantity $l_{n}$ as  
the number of particles satisfying the
inequality $\gamma_{j}^{(n)}>0.2$ just before the $n$-th collision.  
   Note that $ 0\leq \gamma_{j}^{(n)} \leq 1$ by 
definition (\ref{gamma}) of $\gamma_{j}^{(n)}$, so 
$l_{n}$ cannot be larger than $5$. 
   If $l_{n}$ is always 2 as in the low density limit, 
then we can use the hopping distance definition 
(\ref{hopDista0}), but in numerical simulations at finite density 
$l_{n}>2$ can happen as shown 
in Fig. \ref{figX1loczAmpDynGradu}. 
   The problem then is how do we define the 
hopping distance $h^{[n]}$ at the $n$-th collision, 
when $l_{n}>2$.

\begin{figure}[t!]
\vspfigA
\caption{
      Schematic illustrations of the 
   hopping types for the localized region of the 
   Lyapunov vector in quasi-one-dimensional systems 
   at low density. 
      The contours represent the level $\gamma_{j}^{(1)}=0.2$. 
      (These are the contours on the base of  
   Fig. \ref{figX1loczAmpDynGradu}.)  
      The vertical dotted line is  
   collision number $n$ at which the 
   hopping distance of the localized Lyapunov vector is 
   to be determined.  
      The seven illustrations in this figure indicate: 
   (a) the case of $(l_{n},l_{n+1})=(2,2)$, 
   (b1) the case of $(l_{n},l_{n+1})=(2,3)$, 
   (b2) the case of $(l_{n},l_{n+1})=(3,2)$,
   (b3) the case of $(l_{n},l_{n+1})=(3,3)$,
   (c1) the case of $(l_{n},l_{n+1})=(2,4)$,
   (c2) the case of $(l_{n},l_{n+1})=(4,2)$, and 
   (c3) the case of $(l_{n},l_{n+1})=(4,4)$,
   where $l_{n}$ is the number of particles in the localized region 
   of Lyapunov vector just before the $n$-th collision.   
      The numbers in the right-bottom of each illustration 
   give the possible values of the hopping distance.  
   }
\label{figX2hopType}
\vspfigB\end{figure}  
%
   The definitions of the hopping distance for each case 
are categorized as follows.  
\begin{description}
\item[$\langle${\it Case (a): 
$l_{n}=2 \rightarrow l_{n+1}=2$ }$\rangle$] 
   Here only two particle indices are in the 
localized region before and after the collision,
and they are always nearest-neighbors, 
so the hopping distance $h^{[n]}$ is given 
by Eq. (\ref{hopDista0}). 
\item[$\langle${\it Case (b1): 
$l_{n}=2 \rightarrow l_{n+1}=3$} $\rangle$]
  Here we assume that the localized regions 
of the Lyapunov vector are given by 
$\{j_{n},j_{n}+1\}$, $\{j_{n+1},j_{n+1}+1,j_{n+1}+2\}$ 
and $j_{n+1}=j_{n}$ or $j_{n+1}=j_{n}-1$. 
   We take the value of the hopping distance to be 1 (-1) for 
 $h^{[n]}$ where $j_{n+1}=j_{n}$  ($j_{n+1}=j_{n}-1)$. 
\item[$\langle${\it Case (b2): 
$l_{n}=3 \rightarrow l_{n+1}=2$} $\rangle$] 
   In this case we assume that  the localized regions 
of the Lyapunov vector are given by 
$\{j_{n},j_{n}+1, j_{n}+2\}$, $\{j_{n+1},j_{n+1}+1\}$ 
and $j_{n+1}=j_{n}$ or $j_{n+1}=j_{n}+1$, and 
the hopping distance is $0$ in both cases. 
\item[$\langle${\it Case (b3): 
$l_{n}=3 \rightarrow l_{n+1}=3$} $\rangle$]
   In this case we assume that the localized regions 
of the Lyapunov vector are given by 
$\{j_{n},j_{n}+1,j_{n}+2\}$, $\{j_{n+1},j_{n+1}+1,j_{n+1}+2\}$ 
and $j_{n+1}=j_{n}+h^{[n]}$ with the hopping distance $h^{[n]}$. 
   We take into account the case  $h^{[n]}=-1,0$ or $1$ 
only.  
\item[$\langle${\it Case (c1): 
$l_{n}=2 \rightarrow l_{n+1}=4$} $\rangle$] 
   In this case we assume that the localized regions 
of the Lyapunov vector are given by 
$\{j_{n},j_{n}+1\}$, $\{j_{n+1},j_{n+1}+1\}$ 
and $\{j_{n+1}',j_{n+1}'+1\}$ 
satisfying $j_{n+1}'=j_{n}$ and $|j_{n+1}'-j_{n+1}|\geq 2$. 
   The hopping distance is defined by Eq. (\ref{hopDista0}) 
using these $j_{n+1}$ and $j_{n}$. 
%
\item[$\langle${\it Case (c2): 
$l_{n}=4 \rightarrow l_{n+1}=2$} $\rangle$]
   In this case we assume that the localized regions 
of the Lyapunov vector are given by 
$\{j_{n},j_{n}+1\}$, $\{j_{n}',j_{n}'+1\}$ 
and $\{j_{n+1},j_{n+1}+1\}$ 
satisfying $|j_{n}-j_{n}'|\geq 2$ and $|j_{n}-j_{n+1}|\leq 1$.  
   The hopping distance $h^{[n]}$ is  
$h^{[n]}=j_{n+1}-j_{n} = -1,0$ or $1$.
%
\item[$\langle${\it Case (c3): 
$l_{n}=4 \rightarrow l_{n+1}=4$} $\rangle$]
   In this case we consider only the case in which the 
localized regions of Lyapunov vector are given by  
$\{j_{n},j_{n}+1\}$, $\{j_{n}',j_{n}'+1\}$,  
$\{j_{n+1},j_{n+1}+1\}$ and $\{j_{n+1}',j_{n+1}'+1\}$ 
satisfying $j_{n+1}=j_{n}$, $j_{n+1}'=j_{n}'$ 
and $|j_{n}-j_{n}'|\geq 2$.  
   The hopping distance takes the value $0$ 
in this case. 
%
\end{description}
   (For each case above, the corresponding schematic illustration 
is shown in Fig. \ref{figX2hopType}.)
   Here, we use periodic boundary conditions for the 
particle index, so that the localized 
regions $\{N,1\}$, $\{N,1,2\}$ and $\{N-1,N,1\}$ should 
be translated to 
$\{0,1\}$, $\{0,1,2\}$ and $\{-1,0,1\}$, respectively,  
in the above definition of the hopping distance. 
   In the examples shown in Fig. \ref{figX1loczAmpDynGradu},  
the first two hops of the localized Lyapunov vector can be 
described as case (a), 
and the third hop is described as cases (c1) and (c2). 
   Note that asymmetric definitions of hopping distances $h^{[n]}$
between cases (c1) and (c2) [and similarly 
between cases (b1) and (b2)] 
are required so that we can interpret the 
hopping distance of the third non-zero hop 
in Fig. \ref{figX1loczAmpDynGradu} as only $-2$ in spite of it 
involving both cases (c1) and (c2). 
   Using the above hopping distance we count the number $N_{T}(h)$
of hops of distance $h$ in time interval $T$, and  
introduce the normalized hopping rate as $P_{N} (h)/P_{N} (1) \equiv 
\lim_{T\rightarrow\infty} N_{T}(h)/N_{T}(1)$ 
for $h=-[N/2],-[N/2]+1,\cdots,[N/2]$.  
   Notice that there are possibilities apart from those shown above, 
but it is observed that in numerical simulations 
the probabilities of these
are extremely small (for example, more than 96 percent of 
all hops could be categorized this way). 

\begin{figure}[t!]
\vspfigA
\caption{
      Log-log plots of the normalized hopping rates 
   $P_{N} (h)/P_{N} (1)$ 
   for a quasi-one-dimensional system of $50$ hard-disks 
   as a function of the hopping distance $|h|$ 
   at different densities, 
   $d=10^{3}$ (circles), $10^{4}$ (triangles) and 
   $10^{5}$ (squares). 
      The density is a function of $d$ given by   
   $\rho=\pi R^{2}/[(1+d)L_{y}^{2}]$.
      The error bars are given 
   by  $|P_{N} (h)-P_{N} (-h)|/P_{N} (1)$.  
   }
\label{figX3hopRateDensi}
\vspfigB\end{figure}  
%
   In Fig. \ref{figX3hopRateDensi} 
we show the normalized hopping rate $P_{N} (h)/P_{N} (1)$ in 
a quasi-one-dimensional system of $50$ hard-disks 
for $d=10^{3}$ (circles), $10^{4}$ (triangles) and 
$10^{5}$ (squares). 
   The hopping rate is almost 
density-independent in this low density range, 
although it may be very slightly larger  
at the lowest density for $|h|=2,3,\cdots$.

   We also calculate the hopping rate 
for the brick accumulation model 
explained in Sec. \ref{brickmodel} in a similar way. 
   In the brick accumulation model we can introduce 
the localized region of the Lyapunov vectors 
as the site (particle) indices whose 
brick height are highest. 
   For the brick model, cases (b1), (b2), and (b3) 
above cannot happen, and only cases (a), (c1), (c2) and (c3) above 
are taken into account in the numerical calculations. 
   It may be noted that in the brick model, 
case (a) above can happen only when the hopping distance 
is $-1$, $0$ or $1$, 
and Eq. (\ref{hopDista0}) alone  
is not enough to calculate the hopping distance. 
   This is another reason to take into account 
the case where $l_{n}>2$ in the calculation of the hopping distance.

\section{Clock Model for Many-Hard-Particle Systems}
\label{ClockModelManyHardParticleSystems}

   Here we give an extension of the derivation of the clock model 
for the Lyapunov vector dynamics in many-hard-particle systems 
in the limit of low density. 
   We also derive the formula (\ref{largeLyapu}) 
for the largest Lyapunov exponent from the clock model. 
   The one-dimensional version of the clock model is 
the brick accumulation model used in this paper. 
   In particular, we clarify the assumptions needed 
to justify the use of this model 
to discuss the Lyapunov localization. 
   For the basic idea of the clock model, 
see, for example, Refs. \cite{Bei00,Zon02}.   
   However, note that in this appendix we use 
notation and assumptions that are a little different 
from these references, 
so that the derived clock model is consistent 
with the discussions in this paper 
and able to be compared with the numerical results 
of Ref. \cite{Tan03b}.  

   We consider a $\calD$-dimensional system with $N$ hard-disks 
(or hard-spheres, etc.) with identical radius $R$ and mass $M$. 
   We assume that there is no external field in the system 
so that the dynamics is simply 
free-flights, and collisions between two particles. 
   We put $\delta\bfq_{j}$ ($\bfq_{j}$) as the spatial part 
of the Lyapunov vector component (the spatial coordinate) 
of the $j$-th particle, and $\delta\bfp_{j}$ ($\bfp_{j}$) 
as the momentum part of the Lyapunov vector component 
(the momentum) of the $j$-th particle. 


   We take $t=t_{n}$ to be the time of the $n$-th 
collision which involves particles $j_{n}$ 
and $k_{n}$, and define $\tau_{n}$ to be  
\begin{eqnarray}
\tau_{n} \equiv \frac{t_{n} - t_{n-1}}{M},
\label{FreeFlighTime}\end{eqnarray}
so that the $n$-th free flight time is given by $\tau_{n}M$. 
   The free flight part of dynamics 
of the Lyapunov vector is represented as 
\begin{eqnarray}
   \delta\bfq_{j}(t_{n}^{-}) &=& \delta\bfq_{j}(t_{n-1}^{+}) 
      + \tau_{n}\delta\bfp_{j}(t_{n-1}^{+}) ,
      \label{LyaVecFree1}\\
   \delta\bfp_{j}(t_{n}^{-}) &=& \delta\bfp_{j}(t_{n-1}^{+})
\label{LyaVecFree2}\end{eqnarray}
   Here, the argument $t_{n}^{\pm}$ refers to the limit 
of the quantity before the $n$-th collision ($-$) or after the $n$-th
collision ($+$).
   On the other hand,  the change in the Lyapunov vector 
in particle-particle collisions is represented as 
\begin{eqnarray}
   \delta\bfq_{j_{n}}(t_{n}^{+}) 
      &=&  \delta\bfq_{j_{n}}(t_{n}^{-}) 
      + \Theta^{[n]} 
        \delta\bfq_{k_{n}j_{n}}(t_{n}^{-}) ,
      \label{LyaVecCol4} \\
   \delta\bfq_{k_{n}}(t_{n}^{+}) 
      &=&   \delta\bfq_{k_{n}}(t_{n}^{-}) 
      - \Theta^{[n]} 
        \delta\bfq_{k_{n}j_{n}}(t_{n}^{-}) ,
      \label{LyaVecCol5} \\
   \delta\bfq_{l}(t_{n}^{+}) 
      &=& \delta\bfq_{l}(t_{n}^{-}), 
      \;\;\;\mbox{for}\;\; 
         l \in\!\!\!\!\!/ \{ j_{n}, 
         k_{n} \},
      \label{LyaVecCol6} \\
%
   \delta\bfp_{j_{n}}(t_{n}^{+}) 
      &=& \delta\bfp_{j_{n}}(t_{n}^{-}) 
      + \Theta^{[n]} 
      \delta\bfp_{k_{n}j_{n}}(t_{n}^{-})    
      \nonumber \\ && \ls
      +Q^{[n]}\delta\bfq_{k_{n}j_{n}}(t_{n}^{-}) , 
      \label{LyaVecCol1} \\
   \delta\bfp_{k_{n}}(t_{n}^{+}) 
      &=& \delta\bfp_{k_{n}}(t_{n}^{-}) 
      -\Theta^{[n]} 
      \delta\bfp_{k_{n}j_{n}}(t_{n}^{-})     
      \nonumber \\ && \ls
      -Q^{[n]}\delta\bfq_{k_{n}j_{n}}(t_{n}^{-})  ,
      \label{LyaVecCol2} \\
   \delta\bfp_{l}(t_{n}^{+}) 
      &=& \delta\bfp_{l}(t_{n}^{-}), 
      \;\;\;\mbox{for}\;\; 
         l \in\!\!\!\!\!/ \{ j_{n}, 
         k_{n} \},
      \label{LyaVecCol3} 
\end{eqnarray}
where $\delta\bfq_{k_{n}j_{n}} \equiv \delta\bfq_{k_{n}} 
     - \delta\bfq_{j_{n}} $,
$\delta\bfp_{k_{n}j_{n}} \equiv \delta\bfp_{k_{n}} 
      - \delta\bfp_{j_{n}},$ 
%
%
and $\Theta^{[n]}$ and $Q^{[n]}$ are 
defined by 
\begin{eqnarray}
   \Theta^{[n]} &\equiv& 
      \bfsigma^{[n]} \bfsigma^{[n]T},  \\
\label{MatriTheta}
   Q^{[n]} &\equiv& \frac{\bfsigma^{[n]T}\Delta\bfp^{[n]} }{2R}
   \left( I + \frac{\bfsigma^{[n]}\Delta\bfp^{[n]T}}
      {\bfsigma^{[n]T}\Delta\bfp^{[n]}} \right)     
      \nonumber \\ && \ls\ls\ls \times
   \left( I - \frac{\Delta\bfp^{[n]}\bfsigma^{[n]T}}
      {\Delta\bfp^{[n]T}\bfsigma^{[n]}}\right) 
\label{MatriQ}\end{eqnarray}   
with 
\begin{eqnarray}
   \bfsigma^{[n]}
      &\equiv& \frac{\bfq_{k_{n}}(t_{n}^{-}) 
      - \bfq_{j_{n}}(t_{n}^{-})}
         {|\bfq_{k_{n}}(t_{n}^{-}) 
      - \bfq_{j_{n}}(t_{n}^{-})|} ,
      \label{Sigma}\\
   \Delta\bfp^{[n]} &\equiv& 
   \bfp_{k_{n}}(t_{n}^{-}) - \bfp_{j_{n}}(t_{n}^{-})
\label{DeltaP}\end{eqnarray}   
and the $(\calD N)\times(\calD N)$ identity matrix $I$.
   Note that in this appendix we introduce all vectors as 
column vectors, so for example,  
$\bfsigma^{[n]T}\Delta\bfp^{[n]}$ 
is a scalar and $\bfsigma^{[n]}\Delta\bfp^{[n]T}$ is a matrix 
where $T$ is the transpose.
   For later use we note that 
\begin{eqnarray}
   \delta\bfp_{j_{n}}(t_{n}^{+}) 
   + \delta\bfp_{k_{n}}(t_{n}^{+}) 
      = \delta\bfp_{j_{n}}(t_{n}^{-})      
      + \delta\bfp_{k_{n}}(t_{n}^{-})      
\label{conseP}\end{eqnarray}
which can be derived from 
Eqs. (\ref{LyaVecCol1}) and (\ref{LyaVecCol2}).   
   For the derivation of Eqs. 
(\ref{LyaVecCol4}), (\ref{LyaVecCol5}), (\ref{LyaVecCol6}), 
(\ref{LyaVecCol1}), (\ref{LyaVecCol2}) and (\ref{LyaVecCol3})
for the Lyapunov vector dynamics, 
for example, see Ref. \cite{Del96}.


   We consider a low density case, in which 
the free flight time $\tau_{n}M$ is large 
(as the free flight time is inversely proportional 
to the density). 
   This justifies our first approximation for the Lyapunov vector:  
\begin{eqnarray}
   \delta\bfq_{j}(t_{n}^{-}) 
   \sim \tau_{n}\delta\bfp_{j}(t_{n-1}^{+})   
\label{LyaVecFree3}\end{eqnarray}
in the limit of low density, the term containing $\tau_{n}$ 
is much larger than the other term on the right-hand side of 
Eq. (\ref{LyaVecFree1}). 
   This asymptotic relation (\ref{LyaVecFree3}) 
leads to 
\begin{eqnarray}
\delta\bfq_{k_{n}j_{n}}(t_{n}^{-}) \sim 
\tau_{n}\delta\bfp_{k_{n}j_{n}}(t_{n-1}^{+})
\label{LyaVecFree3b}\end{eqnarray}
from the definition of $\delta\bfq_{k_{n}j_{n}}(t_{n}^{-})$ 
and $\delta\bfp_{k_{n}j_{n}}(t_{n-1}^{+})$. 
   Using the relation (\ref{LyaVecFree3b}), 
Eqs. (\ref{LyaVecCol1}), (\ref{LyaVecCol2}) and 
(\ref{LyaVecCol3}) can be rewritten as 
\begin{eqnarray}
   \delta\bfp_{j_{n}}(t_{n}^{+}) 
      &\sim & \delta\bfp_{j_{n}}(t_{n-1}^{+}) 
       +\Theta^{[n]} 
      \delta\bfp_{k_{n}j_{n}}(t_{n-1}^{+})     
      \nonumber \\ && \ls
      +\tau_{n} Q^{[n]}\delta\bfp_{k_{n}j_{n}}(t_{n-1}^{+}) , 
      \label{LyaVecCol1a} \\
   \delta\bfp_{k_{n}}(t_{n}^{+}) 
      &\sim & \delta\bfp_{k_{n}}(t_{n-1}^{+}) 
      -\Theta^{[n]} 
      \delta\bfp_{k_{n}j_{n}}(t_{n-1}^{+})     
      \nonumber \\ && \ls
      -\tau_{n} Q^{[n]}\delta\bfp_{k_{n}j_{n}}(t_{n-1}^{+})  
      \label{LyaVecCol2a}       \\
   \delta\bfp_{l}(t_{n}^{+}) 
      &=& \delta\bfp_{l}(t_{n-1}^{+}), 
      \;\;\;\mbox{for}\;\; 
         l \in\!\!\!\!\!/ \{ j_{n},  k_{n} \},
      \label{LyaVecCol3a} 
\end{eqnarray}
where we have used Eq. (\ref{LyaVecFree2}). 
   Note that the spatial part of the Lyapunov vector does not appear 
in the dynamics described in (\ref{LyaVecCol1a}) and 
(\ref{LyaVecCol2a}), (\ref{LyaVecCol3a}) anymore. 
   The first and second terms on the right-hand side of 
(\ref{LyaVecCol1a}) and (\ref{LyaVecCol2a}) are negligible 
compared with the third term because of the large value of $\tau_{n}$,  
so we obtain 
\begin{eqnarray}
   \delta\bfp_{j_{n}}(t_{n}^{+}) 
      &\sim & \tau_{n} Q^{[n]}\delta\bfp_{k_{n}j_{n}}(t_{n-1}^{+}) ,
      \label{LyaVecCol1b} \\
   \delta\bfp_{k_{n}}(t_{n}^{+}) 
      &\sim & -\tau_{n} Q^{[n]}\delta\bfp_{k_{n}j_{n}}(t_{n-1}^{+}),   
      \label{LyaVecCol2b} 
\end{eqnarray}
which lead to 
\begin{eqnarray}
   \delta\bfp_{j_{n}}(t_{n}^{+}) + \delta\bfp_{k_{n}}(t_{n}^{+}) 
   \sim 0.  
\label{SumPjPk}\end{eqnarray}
   On the other hand, for the dynamics given by 
(\ref{LyaVecCol4}), (\ref{LyaVecCol5}) and (\ref{LyaVecCol6}) 
for the spatial part of Lyapunov vector 
we obtain
\begin{eqnarray}
   \delta\bfq_{j_{n}}(t_{n}^{+}) 
      &\sim &  \tau_{n}\left[\delta\bfp_{j_{n}}(t_{n-1}^{+}) 
      +\Theta^{[n]} 
        \delta\bfp_{k_{n}j_{n}}(t_{n-1}^{+}) \right] ,
      \label{LyaVecCol4b} \\
   \delta\bfq_{k_{n}}(t_{n}^{+}) 
      &\sim &   \tau_{n}\left[\delta\bfp_{k_{n}}(t_{n-1}^{+}) 
      - \Theta^{[n]} 
        \delta\bfp_{k_{n}j_{n}}(t_{n-1}^{+})\right] , \;\;\;\;\;
      \label{LyaVecCol5b} \\
   \delta\bfq_{l}(t_{n}^{+}) 
      &\sim & \tau_{n}\delta\bfp_{l}(t_{n-1}^{+}), 
      \;\;\;\mbox{for}\;\; 
         l \in\!\!\!\!\!/ \{ j_{n},    k_{n} \},
      \label{LyaVecCol6b} 
\end{eqnarray}
using Eqs. (\ref{LyaVecFree3}) and 
(\ref{LyaVecFree3b}). 
   Now we note 
\begin{eqnarray}
   \delta\bfp_{j_{n}}(t_{n-1}^{+}) 
      &=& \frac{\delta\bfp_{j_{n}}(t_{n-1}^{+})
         +\delta\bfp_{k_{n}}(t_{n-1}^{+})}{2} 
      \nonumber \\ && \ls
      + \frac{\delta\bfp_{j_{n}}(t_{n-1}^{+})
         -\delta\bfp_{k_{n}}(t_{n-1}^{+})}{2}
         \nonumber \\
      &=& \frac{\delta\bfp_{j_{n}}(t_{n}^{-})
         +\delta\bfp_{k_{n}}(t_{n}^{-})}{2} 
      + \frac{\delta\bfp_{j_{n}k_{n}}(t_{n-1}^{+})}{2}
         \nonumber \\
      &=& \frac{\delta\bfp_{j_{n}}(t_{n}^{+})
         +\delta\bfp_{k_{n}}(t_{n}^{+})}{2} 
      - \frac{\delta\bfp_{k_{n}j_{n}}(t_{n-1}^{+})}{2}
         \nonumber \\
      &\sim & - \frac{\delta\bfp_{k_{n}j_{n}}(t_{n-1}^{+})}{2}
\end{eqnarray}
where we used Eqs. (\ref{LyaVecFree2}),  
(\ref{conseP}) and (\ref{SumPjPk}). 
   Similarly we have 
\begin{eqnarray}
   \delta\bfp_{k_{n}}(t_{n-1}^{+}) 
      &\sim & \frac{\delta\bfp_{k_{n}j_{n}}(t_{n-1}^{+})}{2}.
\end{eqnarray}
Therefore Eqs. (\ref{LyaVecCol4b})
and (\ref{LyaVecCol5b}) can be rewritten as 
\begin{eqnarray}
   \delta\bfq_{j_{n}}(t_{n}^{+}) 
      &\sim &  \tau_{n} \Omega^{[n]} 
      \delta\bfp_{k_{n}j_{n}}(t_{n-1}^{+}) ,
      \label{LyaVecCol4c} \\
   \delta\bfq_{k_{n}}(t_{n}^{+}) 
      &\sim &   -\tau_{n}\Omega^{[n]}
        \delta\bfp_{k_{n}j_{n}}(t_{n-1}^{+})
      \label{LyaVecCol5c} 
\end{eqnarray}
where $\Omega^{[n]}$ is 
defined by  
\begin{eqnarray}
   \Omega^{[n]} \equiv -\frac{1}{2} I + \Theta^{[n]} .
\end{eqnarray}
   The asymptotic equations (\ref{LyaVecCol3a}),  
(\ref{LyaVecCol1b}), (\ref{LyaVecCol2b}), (\ref{LyaVecCol6b}), 
(\ref{LyaVecCol4c}) and (\ref{LyaVecCol5c}) 
give the Lyapunov vector dynamics in the 
limit of low density.
   It should also be noted that the assumptions used to derive 
this dynamics breaks  
some conservation laws in the original dynamics, 
for example, the quantity $\sum_{j=1}^{N} \delta\bfp_{j}(t)$ is 
conserved in the original dynamics (\ref{LyaVecFree2}), 
(\ref{LyaVecCol1}), (\ref{LyaVecCol2}) and 
(\ref{LyaVecCol3}) but this cannot be guaranteed 
exactly in the low density dynamics 
(\ref{LyaVecCol3a}), (\ref{LyaVecCol1b}) and 
(\ref{LyaVecCol2b}).

   Now we consider the Lyapunov vector $\delta\bfGamma$ 
corresponding to a positive Lyapunov exponent. 
   The positivity of the Lyapunov exponent means that 
the amplitude $|\delta\bfGamma|$ of Lyapunov vector diverges 
exponentially in time. 
   The dynamics given by (\ref{LyaVecCol3a}),  
(\ref{LyaVecCol1b}), (\ref{LyaVecCol2b}), (\ref{LyaVecCol6b}), 
(\ref{LyaVecCol4c}) and (\ref{LyaVecCol5c}) shows that 
the divergence of $|\delta\bfGamma|$ must come from 
the Lyapunov vector components corresponding 
to colliding particles, as the other Lyapunov vector components 
diverge at most linearly.  
   For this reason we neglect the change of the Lyapunov vector 
components for non-colliding particles. 
   Under this assumption the Lyapunov vector dynamics 
for the Lyapunov vector component 
$\delta\bfGamma_{j}\equiv(\delta\bfq_{j},\delta\bfp_{j})^{T}$ 
for the $j$-th particle is summarized as 
\begin{eqnarray}
   \delta\bfGamma_{j_{n}}(t_{n}^{+}) 
      &\sim & - \delta\bfGamma_{k_{n}}(t_{n}^{+}) 
      \sim  \tau_{n} \delta\mathbf{\Xi}^{[n-1]}
      \label{LyaDyn1} \\
   \delta\bfGamma_{l}(t_{n}^{+}) 
      &\sim & \delta\bfGamma_{l}(t_{n-1}^{+}), 
      \;\;\;\mbox{for}\;\; 
         l \in\!\!\!\!\!/ \{ j_{n},   k_{n} \},
      \label{LyaDyn2} 
\end{eqnarray}
where $\delta\mathbf{\Xi}^{[n-1]}$ is 
defined by 
\begin{eqnarray}
   \delta\mathbf{\Xi}^{[n-1]}
   \equiv \left(\begin{array}{c}
  \Omega^{[n]}\delta\bfp_{k_{n}j_{n}}(t_{n-1}^{+}) \\
      Q^{[n]}\delta\bfp_{k_{n}j_{n}}(t_{n-1}^{+})
   \end{array}\right) .
\end{eqnarray}
   It is essential to note that the Lyapunov vector 
components $\delta\bfGamma_{j_{n}}(t_{n}^{+})$ 
and $\delta\bfGamma_{k_{n}}(t_{n}^{+})$ for 
the colliding particles have the same amplitude, 
and $\delta\mathbf{\Xi}^{[n-1]}$ 
is independent of $\tau_{n}$.

   We assume that the ratio 
\begin{eqnarray}
 \mu \equiv \frac{|\delta\bfq_{j}|}{|\delta\bfp_{j}|}
\label{RatioMu} \end{eqnarray}
between the amplitudes 
of the spatial  
and momentum parts of the Lyapunov vector for 
the $j$-th particle, 
are independent of the particle 
index $j$ \cite{noteA}. 
   Ref.  \cite{Tan03b} suggests that the ratio $\mu$ 
need not be of order $1$ in general. 
   From Eq. (\ref{RatioMu}) we have 
\begin{eqnarray} 
   |\delta\bfp_{j}| \approx \frac{1}{\sqrt{1+\mu^{2}}}
   |\delta\bfGamma_{j}|. 
\end{eqnarray}
   Using this assumption we estimate the magnitude 
of the vector $\delta\mathbf{\Xi}^{[n-1]}$ as 
\begin{eqnarray}
   &&\hspace{-0.5cm}\left|\delta\mathbf{\Xi}^{[n-1]}\right| 
   \nonumber \\
   &&= \sqrt{ 
   \left|\Omega^{[n]}\delta\bfp_{k_{n}j_{n}}(t_{n-1}^{+})\right|^{2} 
   + \left| Q^{[n]}\delta\bfp_{k_{n}j_{n}}(t_{n-1}^{+})\right|^{2}}
   \nonumber \\
   &&= \mbox{max}\left\{
         \left|\delta\bfp_{j_{n}}(t_{n-1}^{+})\right|, 
         \left|\delta\bfp_{k_{n}}(t_{n-1}^{+})\right| 
         \right\} 
      \nonumber \\ &&\ls\ls\times
      \sqrt{ 
         \left|\Omega^{[n]}\bfe^{[n]}\right|^{2} 
         + \left| Q^{[n]} \bfe^{[n]}\right|^{2}}
   \nonumber \\
   &&\approx \mbox{max}\left\{
         \left|\delta\bfGamma_{j_{n}}(t_{n-1}^{+})\right|, 
         \left|\delta\bfGamma_{k_{n}}(t_{n-1}^{+})\right| 
         \right\} 
      \nonumber \\ &&\ls\ls\times
      \sqrt{ \frac{
         \left|\Omega^{[n]}\bfe^{[n]}\right|^{2} 
         + \left| Q^{[n]} \bfe^{[n]}\right|^{2}}{1+\mu^{2}}}
\label{magniXi}\end{eqnarray}
where $\bfe^{[n]}$ is defined by 
\begin{eqnarray}
   \bfe^{[n]} 
   \equiv \frac{
   \delta\bfp_{k_{n}}(t_{n-1}^{+}) 
   -\delta\bfp_{j_{n}}(t_{n-1}^{+})}
   {\mbox{max}\left\{
         \left|\delta\bfp_{k_{n}}(t_{n-1}^{+})\right|, 
         \left|\delta\bfp_{j_{n}}(t_{n-1}^{+})\right| 
         \right\} }, 
\label{bfeDef}\end{eqnarray}
and satisfies the inequality 
\begin{eqnarray}
   \left|\bfe^{[n]}\right| 
   \leq \frac{
   \left|\delta\bfp_{k_{n}}(t_{n-1}^{+})\right|  
   +\left|\delta\bfp_{j_{n}}(t_{n-1}^{+})\right| }
   {\mbox{max}\left\{
         \left|\delta\bfp_{k_{n}}(t_{n-1}^{+})\right|, 
         \left|\delta\bfp_{j_{n}}(t_{n-1}^{+})\right| 
         \right\} } \leq 2. \ls
\label{bfeDef2}\end{eqnarray}
   From Eq. (\ref{magniXi}) we can estimate  
the magnitudes $|\delta\bfGamma_{j_{n}}(t_{n}^{+})|$ 
and $|\delta\bfGamma_{k_{n}}(t_{n}^{+})|$
through $|\delta\bfGamma_{j_{n}}(t_{n}^{+})| $ $
= |\delta\bfGamma_{k_{n}}(t_{n}^{+})| = 
\tau_{n}|\delta\mathbf{\Xi}^{[n-1]}|$.

   Now, we introduce the clock value $\mathcal{K}_{j}(n)$ 
of the $j$-th particle just after the $n$-th collision as 
%
\begin{eqnarray}
   \mathcal{K}_{j}(n) 
   \equiv 
   -\frac{1}{\ln\rho}\ln\frac{|\delta\bfGamma_{j}(t_{n}^{+})| }{
   |\delta\bfGamma_{j}(0)|},  
\label{ClockValue1}\end{eqnarray}
or equivalently,  
\begin{eqnarray}
   |\delta\bfGamma_{j}(t_{n}^{+})| 
   = \left(\frac{1}{\rho}\right)^{\mathcal{K}_{j}(n)} 
   |\delta\bfGamma_{j}(0)| 
\label{ClockValue1b}\end{eqnarray}
leading to Eq. (\ref{brikheight}).
   Here $\rho$ is the particle density whose value 
is $0<\rho<1$, and 
$\delta\bfGamma_{j}(0)$ is the Lyapunov vector component 
for the $j$-th particle at the initial time. 
   We choose the initial Lyapunov vector 
$\delta\bfGamma (0)$ so that its components  
$\delta\bfGamma_{1}(0), \delta\bfGamma_{2}(0), \cdots, 
\delta\bfGamma_{N-1}(0)$ and $\delta\bfGamma_{N}(0)$ 
have the same order of magnitude, 
and a larger clock value means larger magnitude 
of the Lyapunov vector component, 
$|\delta\bfGamma_{j}(t_{n})|$ .

   We have already used the fact that the 
free flight time increases as the density $\rho$ decreases.  
   To write down the clock value more meaningfully, 
we use the specific relation between the free flight time 
and the density, namely that the free-flight time $\tau_{n}M$ 
is approximately 
inversely proportional to the density $\rho$ at low density;  
\begin{eqnarray}
   \tau_{n} \sim s_{n}/\rho 
\label{MeanFree}\end{eqnarray}
where $s_{n}$ is independent of the density. 
   Using Eq. (\ref{MeanFree}), the expression (\ref{ClockValue1}) 
for the clock value $\mathcal{K}_{j}(n)$ can be rewritten as 
\begin{eqnarray}
   \mathcal{K}_{l}(n) 
   &\sim& \left\{
      \begin{array}{l}
         \mbox{max}\left\{\mathcal{K}_{j_{n}}(n-1), 
         \mathcal{K}_{k_{n}}(n-1) \right\}+ 1 + \Delta\Phi^{[n]} 
         \\ \ls 
         \;\;\;\mbox{for}\;\; l = j_{n} \;\; 
         \mbox{or}\;\; l = k_{n} \\ \\
         \mathcal{K}_{l}(n-1) 
         \;\;\;\mbox{for}\;\; 
         l \in\!\!\!\!\!/ \{ j_{n},   k_{n} \},
      \end{array}
   \right.
\label{ClockValue2}\end{eqnarray}
where $\Delta\Phi^{[n]}$ is defined by 
\begin{eqnarray}
   \Delta\Phi^{[n]} \equiv 
   -\frac{1}{\ln\rho}\ln \left\{s_{n}\sqrt{ \frac{
         \left|\Omega^{[n]}\bfe^{[n]}\right|^{2} 
         + \left| Q^{[n]} 
         \bfe^{[n]}\right|^{2}}{1+\mu^{2}}}\right\} ,  
   \nonumber \\
\label{DeltaPhi}\end{eqnarray}
and where we have used (\ref{LyaDyn1}), (\ref{magniXi}), 
(\ref{ClockValue1}) and $|\delta\bfGamma_{j_{n}}(0)|
\approx |\delta\bfGamma_{k_{n}}(0)|$.  
   Our final assumption to derive the clock model is that 
\begin{eqnarray}
   \lim_{\rho\rightarrow0} \Delta\Phi^{[n]} = 0.     
\label{assumDeltaPhi}\end{eqnarray}
   To justify the assumption (\ref{assumDeltaPhi}) 
note that the right-hand side of Eq. (\ref{DeltaPhi})  
for the quantity $\Delta\Phi^{[n]}$ has a factor 
$1/\ln\rho$ which goes to zero in the limit as  
$\rho\rightarrow 0$, and 
$s_{n}$, $Q^{[n]}$ 
and $\Omega^{[n]}$ 
are almost independent of the density $\rho$, 
and the magnitude of the vector $\bfe^{[n]}$ is finite 
even in the limit $\rho\rightarrow 0$ from the 
inequality (\ref{bfeDef2})   
\cite{noteA}. 
   Eqs. (\ref{ClockValue2}) and (\ref{assumDeltaPhi}) 
lead to the clock dynamics 
\begin{eqnarray}
   \mathcal{K}_{l}(n) 
   &\sim & \left\{
      \begin{array}{ll}
         \mbox{max}\left\{\mathcal{K}_{j_{n}}(n-1), 
         \mathcal{K}_{k_{n}}(n-1) \right\}+ 1 
         \\ \ls 
         \;\;\;\mbox{for}\;\; l = j_{n} \;\; 
         \mbox{or}\;\; l = k_{n} \\ \\
         \mathcal{K}_{l}(n-1) 
         \;\;\;\mbox{for}\;\; 
         l \in\!\!\!\!\!/ \{ j_{n}, k_{n} \},
      \end{array}
   \right.
\label{ClockValue3}\end{eqnarray}
in the low density limit, 
which is closed by the clock value $\mathcal{K}_{l}(n)$ only. 
   From Eq. (\ref{ClockValue3}), the dynamics for the clock model 
is expressed as 
(i) the clock value is changed only when the corresponding 
particle collides, and (ii) the clock values of colliding particles 
are tuned to the same value given by $1$ plus the larger 
of the two clock values of the particles just before the collision. 

   In the quasi-one-dimensional system with periodic boundary 
conditions, particle indices of the colliding particles can be taken 
so that $k_{n} = j_{n}+1$ (note that index $N+1$ is 
equivalent to $1$). 
   Therefore we obtain the dynamics (\ref{ClockValue4})  
for the brick accumulation model 
explained in Sec. \ref{brickmodel} as the one-dimensional version 
of the clock model. 
   Moreover, in the one-dimensional case, $\mathcal{K}_{j}(n)$ can be 
interpreted as the brick height of the $j$-site just after 
the $n$-th brick is dropped. 

   Finally we derive the formula (\ref{largeLyapu}) 
for the largest Lyapunov exponent 
from the clock value $\mathcal{K}_{j}(n)$. 
   In the limit of low density $\rho << 1$, 
the amplitude $|\delta\bfGamma(t_{n}^{+})|$ of the Lyapunov vector 
can be approximated by 
\begin{eqnarray}
   |\delta\bfGamma(t_{n}^{+})| 
      &=& \sqrt{\sum_{j=1}^{N}
         |\delta\bfGamma_{j}(t_{n}^{+})|^{2} } \nonumber \\
      &\sim& \alpha_{n} 
      \left(\frac{1}{\rho}\right)^{\tilde{\mathcal{K}}^{[max]}(n)}  
      |\delta\bfGamma_{j}(0)| 
\label{AmpLyaVecLowDen1}\end{eqnarray}
noting that the sum $\sum_{j=1}^{N}
|\delta\bfGamma_{j}(t_{n}^{+})|^{2}$ is dominated 
by the Lyapunov vector component amplitude 
$|\delta\bfGamma_{j}(t_{n}^{+})|$ with the largest 
clock value $\tilde{\mathcal{K}}^{[max]}(n)$: 
\begin{eqnarray}
   \tilde{\mathcal{K}}^{[max]}(n) \equiv 
   \max \left\{
      \mathcal{K}_{1}(n), \mathcal{K}_{2}(n), 
      \cdots, \mathcal{K}_{N}(n)
   \right\}
\label{AmpLyaVecLowDen2}\end{eqnarray}
at $n$, because of the huge factor $1/\rho$. 
   Here, $\alpha_{n}$ is the number of particle 
with the largest clock value $\tilde{\mathcal{K}}^{[max]}(n)$, 
and we have also used our assumption that 
$|\delta\bfGamma_{j}(0)|$ is almost independent of the 
particle index $j$. 
   Now we assume the approximate relation 
\begin{eqnarray}
   \frac{\tilde{\mathcal{K}}^{[max]}(n) }{n}
   \;\stackrel{n\rightarrow\infty}{\sim}\;
   \frac{1}{nN}\sum_{j=1}^{N}
   \mathcal{K}_{j}(n)
\label{AmpLyaVecLowDen3}\end{eqnarray}
in the limit of large $n$. 
   (We have checked this relation numerically 
for the brick accumulation model.)
   Using the relations (\ref{AmpLyaVecLowDen1}), 
(\ref{AmpLyaVecLowDen3}) and $t\sim n\tau$ we obtain 

\begin{eqnarray}
   \lambda &\equiv& \lim_{t\rightarrow+\infty} 
   \frac{1}{t}\ln |\delta\bfGamma(t)| \nonumber \\ 
   &\sim& - \lim_{n\rightarrow+\infty} \frac{1}{n\tau} 
   \tilde{\mathcal{K}}^{[max]}(n) \ln\rho \nonumber \\ 
   &\sim& - \lim_{n\rightarrow+\infty} \frac{1}{n\tau N} 
   \sum_{j=1}^{N}
   \mathcal{K}_{j}(n)\ln\rho
\label{largeLyaExpo3}\end{eqnarray}
for the Lyapunov exponent $\lambda$ corresponding to the 
Lyapunov vector $\delta\bfGamma$. 
   Therefore we obtain the Eq. (\ref{largeLyapu}), 
which is independent of the system shape and the 
number of spatial dimensions.




\begin{thebibliography}{99}
\bibitem{noteA} 
   Ref. \cite{Tan03b} suggests that the ratio $\mu$ 
depends on the density $\rho$, and it can be $\mu >> 1$ 
for a Lyapunov vector corresponding 
to a positive Lyapunov exponent at low density.  
   Therefore the assumption 
(\ref{assumDeltaPhi}) is not obvious.
\bibitem{noteSum} 
   In the actual simulation of the brick accumulation model 
we discarded the brick accumulations after the first $100N$ 
bricks have been dropped. 
   In this sense, the sum 
$\sum_{j=1}^{N} \mathcal{K}_{j}(n)$ is not zero at $n=0$. 
   This is the reason why we added the fitting parameter 
$\alpha$ in the fitting function 
$\sum_{j=1}^{N} \mathcal{K}_{j}(n) =\alpha + \beta n$ 
in Fig. \ref{figC3brickSum} 
for consistency, although it is almost zero in this case.
\bibitem{noteB} 
   More exactly, the distribution 
$D(\bfsigma\cdot\Delta\bfp)$ for non-zero hopping cases 
in Fig. \ref{figE1ndpdistri} was calculated as follows. 
   First we chose the case satisfying the conditions: 
(i) the particle indices satisfying 
$\gamma_{j}^{(1)}>0.2$ are given by $\{j_{n},j_{n}+1\}$ 
just before the $n$-th collision and 
$\{j_{n+1},j_{n+1}+1\}$ just after the $n$-th collision, 
such that $j_{n}\neq j_{n+1}$, and 
(ii) the $j_{n+1}$-th particle and 
$(j_{n+1}+1)$-th particle collide at the $n$-th collision.
   (Here, if the particle indices for the localized region of the 
Lyapunov vector satisfy $\gamma_{j}^{(1)}>0.2$ is  $\{1,N\}$, 
then the particle indices pair 
$\{j_{n},j_{n}+1\}$ or $\{j_{n+1},j_{n+1}+1\}$ above 
should be taken as $\{0,1\}$, based on periodic boundary conditions 
in the longitudinal direction of the quasi-one-dimensional system.)  
   Under this circumstance we calculate the distribution 
$D(\bfsigma\cdot\Delta\bfp)$ from the value of 
$\bfsigma\cdot\Delta\bfp$ in the $n$-th collision 
between these $j_{n+1}$-th and $(j_{n+1}+1)$-th particles.
\end{thebibliography}
\end{document}